\documentclass[a4paper]{jpp}
\usepackage{graphicx}
\usepackage[T1]{fontenc}
\usepackage{amsmath}
\usepackage{bm}
\usepackage{natbib}
\usepackage{float}
\usepackage[colorlinks=true,
  linkcolor=black, citecolor=blue, urlcolor=blue]{hyperref}   
\usepackage[dvipsnames,svgnames,x11names,hyperref]{xcolor}  
\usepackage[capitalize]{cleveref}
\usepackage{parskip}

\usepackage{setspace}
\crefformat{subsection}{\S#1}

\shorttitle{The hot-electron closure of the moment-based gyrokinetic plasma model}
\shortauthor{A C.D. Hoffmann et al.}

\title{The hot-electron closure of the moment-based gyrokinetic plasma model}
\author{A.C.D. Hoffmann \aff{1}\corresp{\email{ahoffmann@pppl.gov}}, P. Giroud-Garampon \aff{1}, P. Ricci \aff{1}}
\affiliation{\aff{1} Ecole Polytechnique F\'ed\'erale de Lausanne (EPFL), Swiss
Plasma Center, CH-1015 Lausanne, Switzerland
}
 \newcommand{\squareparenthesis}[1]{\left[#1\right]} 
 \newcommand{\roundparenthesis}[1]{\left(#1\right)} 

 \newcommand{\PJ}[2]{$(P,J)=(#1,#2)$}

 \newcommand{\lperp}{\ell_\perp}

 \newcommand{\spar}{s_{\parallel}}
 
 \newcommand{\wperp}{w_{\perp}}
 
 \newcommand{\upar}{u_{\parallel}}

 \newcommand{\Tperp}{T_{\perp}}
 
 \newcommand{\Tpar}{T_{\parallel }}

 \newcommand{\Cpar}{\hat C_{\parallel}}
 \newcommand{\Cper}{\hat C_{\perp}}

 \newcommand{\ddt}{\partial_t}

 \newcommand{\ddy}{\partial_y}

\newcommand{\kernel}{\hat K}

\newcommand{\ExB}{\bm E\times\bm B}
\newcommand{\order}[1]{\mathcal O\left(#1\right)}
\newcommand{\pb}[2]{\left\{#1,#2\right\}}

\newcommand{\gyacomo}{\textsc{Gyacomo} }

\newcommand{\modifi}[1]{{#1}}

\usepackage{natbib}
\usepackage{graphicx}
\usepackage{bm}
\usepackage{cancel}
\usepackage{float}
\usepackage{algorithm}
\usepackage{algpseudocode}
\usepackage{multicol}

\begin{document}

\maketitle

\begin{abstract}
We derive the hot-electron-limit (HEL) closure for the moment hierarchy used to solve the gyrokinetic equations, known as the gyromoment (GM) approach. By expanding the gyroaveraging kernels in the small temperature ratio limit, $\tau = T_i/T_e \ll 1$, and retaining only the essential $\mathcal{O}(\tau)$ terms, we obtain a closed system for the density, parallel velocity, and parallel and perpendicular temperatures. In a Z-pinch geometry, the GM system with the HEL closure is analytically equivalent to the one developed by \cite{Ivanov2022DimitsTurbulence}.
Numerical benchmarks confirm the closure's accuracy, reproducing established linear growth rates, nonlinear heat transport, and low collisionality dynamics. An extension to the tokamak-relevant $s{-}\alpha$ geometry and a comparison with gyrokinetic simulations reveal the capabilities and limitations of the HEL-closed GM model: while transport levels and temporal dynamics are qualitatively preserved even at $\tau=1$, the absence of higher-order kinetic moments prevents an accurate prediction of the Dimits shift and of transport suppression.
\end{abstract}

\section{Introduction}

A predictive understanding of turbulent transport in magnetized fusion plasmas relies largely on first-principles gyrokinetic (GK) simulations, whose five-dimensional phase-space resolution is computationally demanding. This computational cost limits, for example, the rapid exploration of the parameter space of fusion devices, ultimately hindering their development.
Moment-based formulations of the GK model \citep{Jorge2017ACollisionality,Mandell2018,Frei2020,frei2025turbulencetransportspectrallyaccelerated} recast the velocity-space dependence into a hierarchy of coupled Hermite-Laguerre moments, which we refer to as \textit{gyromoments} (GMs). In practice, simulation and theoretical results show that substantially fewer moments than grid points are often needed for comparable accuracy \citep{Frei2022LocalMode,Hoffmann2023GyrokineticOperators,frei2025turbulencetransportspectrallyaccelerated}.

The GM hierarchy requires a closure because the evolution equation for each moment is coupled to higher-order moments, a consequence of phase-mixing and finite-Larmor-radius (FLR) effects \citep{Grant1967Fourier-HermiteLimit,Jorge2017ACollisionality}. 
\modifi{Closure schemes for moment hierarchies have a long history in plasma turbulence research. Earlier closures of gyrofluid models \citep{Hammett1990FluidInstability,Dorland1993GyrofluidEffects,Waltz1997AModel,Snyder1997LandauFluidMHD}, despite showing very close agreement with full GK simulations in linear growth rates and Landau damping, often struggled to accurately capture nonlinear turbulence saturation levels and zonal-flow (ZF) dynamics \citep{Dimits2000ComparisonsSimulations}. 
More recently, a naive truncation closure, i.e., setting all moments above a certain cutoff to zero, showed very good agreement once the number of retained moments was sufficiently large \citep{Hoffmann2023GyrokineticShift,Frei2023Moment-basedModel}. However, when the number of moments retained is comparable to that of gyrofluid models, such a truncation breaks conservation properties, leading to erroneous transport predictions. 
At the same time, approaches inspired by large-eddy-simulation techniques have been explored in gyrokinetics, using truncation and/or selective damping of higher-order moments to reduce computational cost while maintaining physical fidelity \citep{MorelLargeEddySimulations,Navarro2014ApplicationsGyrokineticTurbulence}. 
Despite these advances, an effective and systematically justified closure that preserves nonlinear regulation mechanisms when only a reduced number of GMs is retained to minimize computational cost remains an open issue and motivates the present work.}

Asymptotic limits provide one possible route to formulating a closure for the GM model. Applying the hot-electron limit (HEL), that is, considering $\tau = T_i/T_e \ll 1$, to the local $\delta f$ GK system \citep{Beer1995}, with $\mathcal{O}(\tau)$ corrections retained only where needed to preserve leading dynamical couplings, yields a three-field fluid model for density, parallel velocity, and parallel temperature, as derived by \citet{Ivanov2020ZonallyTurbulence,Ivanov2022DimitsTurbulence}. We refer to this model as the \emph{Ivanov model}, which is shown to successfully reproduce a Dimits shift and ZF features in a Z-pinch geometry (with adiabatic electrons).
Despite these promising results in a simplified geometry, to our knowledge, the Ivanov model has not yet been systematically benchmarked against GK simulations, leaving the scope and limits of its validity uncertain. This calls for further investigation of the HEL closure to understand, for instance, whether the Ivanov model appears within the GM hierarchy under the same ordering and whether a GK code can reproduce HEL results using a suitably small $\tau$ parameter. In addition, the application of the closure to tokamak geometry and the comparison with finite-$\tau$ GK results remain open issues.

In this paper, we address these points by deriving the HEL of the GM model through a proper expansion in $\tau$. A set of fluid equations (density, parallel velocity, parallel temperature, and perpendicular temperature) is obtained. We show analytically the equivalence of this model with the Ivanov model in a Z-pinch geometry, and we confirm numerically that the GM hierarchy yields a closed four-moment system that reproduces the linear and nonlinear results of the Ivanov model to good accuracy in the small-$\tau$ regime.
In particular, our numerical study benchmarks linear Z-pinch ITG growth-rate convergence against previous results \citep{Ivanov2020ZonallyTurbulence,Ivanov2022DimitsTurbulence}. Nonlinear simulations accurately recover heat flux levels and reproduce bursty or blow-up behavior at low collisionality, capturing the transition where ZF reinforcement weakens, as also shown by \cite{Ivanov2020ZonallyTurbulence,Ivanov2022DimitsTurbulence}. A detailed analysis of turbulence prediction in the Z-pinch geometry is then presented. Next, we extend the HEL GM model to more complex geometries.
Specifically, we focus on the tokamak $s{-}\alpha$ geometry for Cyclone Base Case (CBC) parameters \citep{Lin1999EffectsTransport}, a standard test case considered by many GK codes \citep{Dimits2000ComparisonsSimulations}.
The HEL GM model predicts an ITG-like instability and qualitatively accurate heat flux levels in comparison with GK simulations.
This indicates that the HEL closure can be applied outside its formal range of validity.
On the other hand, the HEL GM model overpredicts transport when approaching marginal stability.
For instance, we do not observe a Dimits shift in the HEL model when considering the tokamak geometry, which indicates that the HEL closure does not overcome the limitations of the lowest-order moment truncation observed in \cite{Hoffmann2023GyrokineticShift}.
This shortfall indicates that higher-order moments beyond the four retained in the HEL closure play an essential role in sustaining zonal-flow regulation closer to marginal stability. This comparison also highlights the more favorable conditions for zonal-flow activity in a Z-pinch geometry.

The numerical results presented here are obtained with \gyacomo \citep{Hoffmann2023GyrokineticShift}, a numerical simulation code that solves the local $\delta f$ GK equation using the GM approach. The code uses field-aligned coordinates and a Fourier representation in the perpendicular plane, allowing for efficient simulations of plasma turbulence in both Z-pinch and tokamak geometries. The HEL GM model is implemented in \gyacomo by retaining only the four lowest-order moments and scaling the gradients and collisionality according to the HEL ordering.

The remainder of this paper is organized as follows. Section~\ref{sec:HEL_GM_model} presents the GM hierarchy, its HEL closure, and its simplification in the Z-pinch geometry. Section~\ref{sec:linear_results} reports benchmarks against existing results. Section~\ref{sec:nonlinear_results} examines nonlinear Z-pinch turbulence in two and three dimensions. Section~\ref{sec:HEL_CBC} extends the application of the HEL closure to the $s$-$\alpha$ geometry and finite $\tau$. Finally, Section~\ref{sec:conclusions} summarizes our findings and outlines possible extensions of the moment closure.

\section{Gyrokinetic model}
\label{sec:HEL_GM_model}
We model ion-scale turbulence in a magnetized plasma using the local, electrostatic, $\delta f$ GK framework with an adiabatic electron response \citep{Catto1978LinearizedGyro-kinetics}.
The model evolves the perturbed ion distribution function $g_i(x,y,z,\spar,\wperp,t)$ in field-aligned coordinates \citep{Beer1995}, where $x$ represents the direction perpendicular to the magnetic flux surface, $y$ the field-line label, $z$ the coordinate aligned with the magnetic field, $\spar$ the velocity parallel to the magnetic field, $\wperp$ the magnetic moment, and $t$ the time. The perturbed distribution function satisfies the gyrokinetic equation, which, in normalized units (see Tab. \ref{tab:dimensionless_units}), is given by \cite{Frei2022LocalMode,Hoffmann2023GyrokineticShift},
\begin{align}
    \ddt g_i 
    &
    +  \{\langle\phi\rangle, g_i\}_{xy}
    + \sqrt{2\tau} s_\parallel \hat C_\parallel h_i - \frac{\sqrt{2}}{2}\sqrt{\tau} w_\perp\hat C_\parallel \ln B \partial_{s_\parallel} h_i
    + \frac{\tau }{q_i}\roundparenthesis{2 s_\parallel^2+w_\perp } \hat C_{\perp} h_i 
    \nonumber \\
    &
    +\squareparenthesis{R_N + \roundparenthesis{s_\parallel^2+w_\perp-\frac{3}{2}} R_T}\ddy \langle\phi\rangle
    = C_{i},
    \label{eq:3d_gyboeq_nondim}
\end{align}

\begin{table}
    \centering
    \begin{tabular}{l r l| l r l}
    Parallel velocity     & $\spar$ & $= v_{\parallel}^{ph}/v_{th i}$  
    &
    \modifi{Magnetic moment}     & $\wperp$ & $= \mu^{ph} B_0/T_{i0}$ 
    \\
    Wave numbers & $k_{x,y}$ & $= k_{x,y}^{ph}\modifi{\rho_{s}}$
    &
    Normalized time & $t$ & $= t^{ph} \modifi{c_{s}}/R_0$
    \\
    Density gradient & $R_{N}$ & $= R_0/L_{N}$ 
    &
    Temperature gradient & $R_{T}$ & $= R_0/L_{T}$ 
    \\
    Ion charge & $q_i$ & $= q_i^{ph}/e$ 
    &
    Temperature ratio & $\tau$ & $= T_{i0}/T_{e0}$
    \\
    Electrostat. potential & $\phi$ & $= e\phi^{ph}/T_{i0} $
    &
    Distribution function & $g_i$ & $= g^{ph}_i/F_{i0}$ \\
    Magnetic field & $B$ & $= B^{ph}/B_0$
    &
    Collision frequency & $\nu$ & $= \nu^{ph} R_0/\modifi{c_{s}}$
    \\
    \end{tabular}
    \caption{Dimensionless variables used throughout the paper. 
    For a dimensionless variable $A$, its equivalent in physical units is explicitly denoted as $A^{ph}$.
    \modifi{We introduce the sound velocity $c_{s}=\sqrt{T_{e0}/m_s}$, the ion thermal velocity $v_{th i} = \sqrt{T_{i0}/m_i}$, the magnetic moment $\mu$}, the reference electron temperature $T_{e0}$, the reference ion temperature $T_{i0}$, the ion thermal Larmor radius $\rho_s = c_{s}/\Omega_i$ with $\Omega_i = q_i^{ph} B_0/m_i$ the ion cyclotron frequency, the reference length scale $R_0$, the reference magnetic field $B_0$, the density and temperature gradient length scales $L_N$ and $L_T$, respectively, and the equilibrium Maxwellian distribution function $F_{i0}$.}
    \label{tab:dimensionless_units}
\end{table}

In Eq. \eqref{eq:3d_gyboeq_nondim}, we introduce the gyroaveraged electrostatic potential $\langle\phi\rangle$, the Jacobian of the field-aligned coordinate system $J_{xyz}$, and the non-adiabatic part of the normalized ion distribution function, $h_i = g_i - \langle\phi\rangle$. The parameter $\tau = T_i/T_e$ denotes the temperature ratio, while $R_N$ and $R_T$ represent the density and temperature gradient parameters, respectively.
The Poisson bracket $\{f_1,f_2\}_{xy}=\partial_x f_1 \partial_y f_2 - \partial_y f_1 \partial_x f_2$ arises from the nonlinear $\ExB$ drift term, while 
\modifi{
\begin{equation}
    \hat C_\parallel = \frac{R_0}{J_{xyz}\hat B}\frac{\partial}{\partial z}
\end{equation}
denotes the magnetic parallel operator, with $R_0$ being a reference length scale, and
\begin{equation}
    \hat C_{\perp}= -\langle[\partial_y \ln B + \frac{G_2}{G_1}\partial_z \ln B \rangle]\frac{\partial}{\partial x}
    + \langle[\partial_x \ln B - \frac{G_3}{G_1}\partial_z \ln B \rangle]\frac{\partial}{\partial y}
\end{equation}
is the magnetic perpendicular operator, where $G_1=g^{xx}g^{yy} - (g^{xy})^2$, $G_2=g^{xx}g^{yz} - g^{xy}g^{xz}$, and $G_3=g^{xy}g^{yz} - g^{yy}g^{xz}$, and where $g^{ij}=\nabla i \cdot \nabla j$ are the metric coefficients for $i,j \in \{x,y,z\}$ \citep{Dhaeseleer1991FluxStructure,Beer1995,Frei2022LocalMode}.} The collision term is represented by $C_i$.
The electrostatic potential is determined from the quasi-neutrality condition, assuming an adiabatic electron response,
\begin{equation}
     \left(1 + \frac{q_i}{\tau}\left[ 1 - \Gamma_0 \right]\right)\phi = q_i n_i + \bar\phi_{yz},
    \label{eq:poisson_adiabe}
\end{equation}
where $\Gamma_0 = I_0(\rho_i^2\nabla_\perp^2)e^{-\rho_i^2\nabla_\perp^2}$, with $I_0$ being the modified Bessel function of the first kind, $\rho_i$ the ion Larmor radius, $\nabla_\perp$ the perpendicular Laplacian, $n_i$ the ion density fluctuation, and $\bar\phi_{yz}$ the potential averaged over the $y$ and $z$ directions.
Details of the GK model can be found in \cite{Hoffmann2023GyrokineticShift}. 

\subsection{Moment-based approach}
To solve the GK Boltzmann equation, we adopt a moment-based approach, projecting the ion GK distribution function onto a basis of Fourier--Hermite--Laguerre modes \citep{Mandell2018,Hoffmann2023GyrokineticOperators,Frei2023Moment-basedModel,Hoffmann2023GyrokineticShift,Mandell2023GX:Design}.
We denote these modes as gyromoments (GMs) and express them as
\begin{equation}
    N_i^{pj}(k_x,k_y,z,t)  =  \iint \mathrm d x \,\mathrm d y  \iint \mathrm d\wperp \,\mathrm d\spar   g_i \,H_p(\spar)\,L_j(\wperp)\, e^{-i(k_x x + k_y y)},
    \label{eq:GMs}
\end{equation}
where $k_x$ is the radial wave number, $k_y$ the binormal wave number, $H_p$ the normalized physicist's Hermite polynomial of order $p$, and $L_j$ the Laguerre polynomial of order $j$.  

In this framework, the gyro-averaging operator can be expressed in terms of Laguerre polynomials as
\begin{equation}
    \langle \phi \rangle =  \sum_{n=0}^\infty \kernel_n(\lperp)\,L_n(\wperp)\phi,
\label{eq:bess_lag}
\end{equation}
where $\lperp = k_\perp^2/2$ and $k_\perp^2 = g^{xx}k_x^2 \,+\,2\, g^{xy}k_xk_y \,+\, g^{yy}k_y^2$. The functions
\begin{equation}
    \kernel_n(\lperp)  =  \frac{(\tau\lperp)^n}{n!} \, e^{-\tau\lperp}
    \label{eq:kernel}
\end{equation}
serve as kernels that separate the configuration- and velocity-space dependencies.  

By projecting the local $\delta f$ GK Boltzmann equation onto the Hermite--Laguerre basis, one obtains the following set of GM equations \citep{Hoffmann2023GyrokineticShift},
\begin{equation}
    \ddt N_i^{pj}
     + \mathcal S^{pj}
     + \mathcal M_{\parallel }^{pj}
     + \mathcal M_{\perp}^{pj}
     + \mathcal D_{T}^{pj}
     + \mathcal D_{N}^{pj}
     = \mathcal C_i^{pj}.
    \label{eq:moment_hierarchy}
\end{equation}
In Eq.~\eqref{eq:moment_hierarchy}, the nonlinear $\ExB$ drift term \modifi{is}, 
\begin{equation}
    \mathcal S^{pj} = \sum_{n=0}^{\infty}\pb{\kernel_i^n\phi}{\sum_{s=0}^{n+j}d_{njs} N_i^{ps}}_{k_x,k_y}\label{eq:exbdrift},
\end{equation}
where the Poisson bracket in Fourier space, $\pb{\cdot}{\cdot}_{k_x,k_y}$, and the Laguerre convolution coefficients, $d_{njs}$, such that $L_nL_j=\sum_{s=0}^{n+j}d_{njs}L_s$ \citep{Gillis1960ProductsPolynomials}. 
The trapping and Landau damping term \modifi{is}, 
\begin{align}
    \mathcal M_{\parallel }^{pj} = &\sqrt{\tau} \left(\Cpar\aleph_i^{p\pm1,j} - C_\parallel^B \left[(j+1)\aleph_i^{p\pm1,j}-j\aleph_i^{p\pm1,j-1}\right]\right)\nonumber\\
    &+\sqrt{\tau}C_\parallel^{B}\sqrt{p}\left([2j+1]n_i^{p-1,j} -[j+1]n_i^{p-1,j+1} - j n_i^{p-1,j-1}\right)\label{eq:landdamp},
\end{align}
with $\aleph_i^{p\pm1,j}=\sqrt{p+1} n_i^{p+1,j} + \sqrt{p} n_i^{p-1,j}$ defined in terms of the non-adiabatic GMs, 
\begin{equation}
    n_i^{pj}=N_i^{pj}+q_i/\tau \kernel_i^j\phi\delta_{p0}. \label{eq:nonadiabatic_GMs}
\end{equation}
The magnetic centrifugal and perpendicular gradient drift term \modifi{is},
\begin{align}
    \mathcal M_{\perp}^{pj} = &\frac{\tau}{q_i} \Cper \squareparenthesis{\sqrt{(p+1)(p+2)} n_i^{p+2,j} + (2p+1)n_i^{pj} + \sqrt{p(p-1)}n_i^{p-2,j}}\label{eq:centforce}\\
    &+ \frac{\tau}{q_i} \Cper \squareparenthesis{(2j+1)n_i^{pj} - (j+1)n_i^{p,j+1}-jn_i^{p,j-1}},\label{eq:magperp}
\end{align}
while the diamagnetic temperature and density gradient drift terms are given by,
\begin{equation}
    \mathcal D_T^{pj} = R_T ik_y\left(\kernel_i^j\squareparenthesis{\frac{1}{\sqrt{2}}\delta_{p2} -\delta_{p0}}+ \squareparenthesis{(2j+1)\kernel_i^j-(j+1)\kernel_i^{j+1}-j\kernel_i^{j-1}}\delta_{p0}\right)\phi,\label{eq:diatemp}
\end{equation}
and
\begin{equation}
     \mathcal D_N^{pj} = R_N ik_y \kernel_a^j \phi\delta_{p0},
    \label{eq:diadens}
\end{equation}
respectively. Finally, $\mathcal C_i^{pj}$ denotes the projection of the ion--ion collision term.  

When considering an adiabatic electron response, the GM equations are closed by the quasi-neutrality relation,
\begin{equation}
     \Bigl(1 + \frac{q_i^2}{\tau}\bigl[1-\!\!\sum_{n=0}^{\infty}\kernel_n^2\bigr]\Bigr)\,\phi  - \bar\phi_{yz}
      =  q_i\,\sum_{n=0}^{\infty}\kernel_n\, N_i^{0n},
    \label{eq:poisson_moments_adiabe}
\end{equation}
where the relation $\Gamma_0 = \sum_{n=0}^{\infty}\kernel_n^2$ is used \citep{Frei2020}.

We refer to the system of Eqs. \eqref{eq:moment_hierarchy} and \eqref{eq:poisson_moments_adiabe} as the \emph{GM model}. It describes the evolution of the GMs, $N_i^{pj}$, and is equivalent to the local GK model in the limit of an infinite number of GMs being retained ($p,j\to\infty$).

To solve the GM system, we use the \gyacomo code\footnote{The version of \gyacomo used in this work is the commit \texttt{fbb6b65b} of the open-source Git repository \texttt{gitlab.epfl.ch/ahoffman/gyacomo}.} \citep{Hoffmann2023GyrokineticOperators,Hoffmann2023GyrokineticShift,Hoffmann2025InvestigationSimulations}, which uses a Fourier approach for the spatial directions and a fourth-order explicit Runge-Kutta scheme for time integration.
The nonlinear term is treated with a $2/3$-dealiasing method \citep{Orszag1971}. 
The evolved Fourier modes have perpendicular wave numbers $k_x = 2\pi m N_x/L_x$ and $k_y = 2\pi n N_y/L_y$, for $m = -N_x/2+1,\ldots,N_x/2$, and $n = 1\ldots,N_y/2$, where $N_x$ and $N_y$ represent the radial and binormal resolutions, and $L_x$ and $L_y$ are the radial and binormal box lengths, respectively.
A hyperdiffusion damping term of the form $\mu_{hd} (k_\perp/k_{\perp,max})^4 N_i^{pj}$, where $\mu_{hd}$ is the hyperdiffusion parameter and $k_{\perp,max}$ is the maximum perpendicular wave number in the simulation, is added in the perpendicular direction to dissipate high-frequency modes, as is often done in nonlinear simulations \citep{Jenko2000,Hoffmann2023GyrokineticOperators}.
Derivatives along the parallel direction, which has length $L_z$, are discretized using a second-order finite-difference scheme on a uniform grid with resolution $N_z$.
\gyacomo supports both tokamak and Z-pinch geometries: in the tokamak geometry, twist-and-shift periodic parallel boundary conditions are applied \citep{Beer1995}, while in the Z-pinch geometry, standard periodic parallel boundary conditions are used \citep{Hoffmann2023GyrokineticOperators}.
For the tokamak geometry, the flux tube length is given by $L_z = 2\pi N_{pol} R_0$, where $R_0$ is the magnetic axis radius and $N_{pol}$ is the number of poloidal turns.
In the Z-pinch geometry, $L_z = 2\pi L_B$, with $L_B$ as the reference magnetic field length scale, corresponding to the major radius in the tokamak geometry or the pinch radius in the Z-pinch geometry.
For the purpose of comparing the two geometries, we set $R_0 = L_B$ so that $N_{pol}$ can be interpreted as the number of times the field line wraps around the Z-pinch axis.
In previous publications \citep{Hoffmann2023GyrokineticOperators,Hoffmann2023GyrokineticShift,Hoffmann2025InvestigationSimulations}, the \gyacomo code evolves the GM model using a truncation closure, setting all GMs with $(p,j)>(p_{\max},j_{\max})$, \modifi{where $p_{\max}$ and $j_{\max}$ are the maximum degrees considered for the Hermite--Laguerre basis. 
In this work, we use a single maximal-degree truncation closure, i.e., we set all GMs with $p+2j > d_{\max}$ to vanish, where $d_{\max}$ is the maximum moment degree considered. This truncation reflects the structure of fluid models, where the moments are evolved up to a certain velocity-polynomial degree, e.g., $d_{\max}=2$ for a Braginskii-type model \citep{Braginskii1965}.}

\subsection{Hot-electron closure}
\label{sec:HEL_closure}
In this subsection, we consider the GM hierarchy up to order $d_{\max}=2$.
This corresponds to evolving the four GMs $N_i^{00}$, $N_i^{10}$, $N_i^{20}$, and $N_i^{01}$ in the limit $\tau\ll 1$ for a single-charge ion species ($q_i=1$). \modifi{In the following, we first write our equations conserving all terms up to a certain power of $\tau$, without assessing whether these terms are dominant or negligible. The ordering of the different terms will be discussed in the next subsection, where the equivalence with the Ivanov model is established.}
We expand the kernel functions $\kernel_n$, Eq.~\eqref{eq:kernel}, for small $\tau$, namely,
\begin{align}
    \kernel_0  &=  1  - \lperp\,\tau 
                  + \tfrac12\,\lperp^2\,\tau^2  + \order{\tau^3},\\
    \kernel_1  &=  \lperp\,\tau 
                  - \lperp^2\,\tau^2 
                  + \order{\tau^3},\\    
    \kernel_2  &=  \tfrac12\,\lperp^2\,\tau^2  + \order{\tau^3}.
\end{align}
The non-adiabatic parts of the ion GMs, Eq. \eqref{eq:nonadiabatic_GMs}, can be written for $(p,j)=(0,0),\,(0,1),\,(0,2)$ as
\begin{align}
    n_i^{00}  &=  N_i^{00}  + \Bigl(\tau^{-1}-\lperp + \tfrac12\,\lperp^2\,\tau\Bigr)\,\phi  + \order{\tau^2},\\
    n_i^{01}  &=  N_i^{01}  + \bigl(\lperp - \lperp^2\,\tau\bigr)\,\phi  + \order{\tau^2},\\
    n_i^{02}  &=  N_i^{02}  + \tfrac12\,\lperp^2\,\tau \,\phi  + \order{\tau^2},
\end{align}
while $n_i^{p,j} = N_i^{p,j} + \order{\tau^2}$ for all $p>0$ and $j>2$. 
\modifi{We note that one may be concerned that the expansion of $n_i^{00}$ has a $\tau^{-1}$ term, which would imply that $n_i^{00}$ scales as $\mathcal{O}(\tau^{-1})$. However, this will be canceled out by the renormalization of $\phi$ when the equivalence with the Ivanov model is established in the next subsection.}

Next, we identify the low-order GMs as \emph{pseudo-fluid moments},
\begin{equation*}
    n^*=N_i^{00},\quad
    \upar^*=N_i^{10},\quad
    \Tpar^*=N_i^{20},\quad
    \Tperp^*=N_i^{01},\quad
    q_\parallel^*=N_i^{30},\quad
    q_\perp^*=N_i^{11},
\end{equation*}
and
\begin{equation*}
    P_\parallel^{\parallel*}=N_i^{40},\quad
    P_\parallel^{\perp*}=N_i^{21},\quad
    P_\perp^{\perp*}=N_i^{02}.
\end{equation*}
These differ slightly from standard fluid moments because the Hermite--Laguerre basis does not match the usual polynomial basis used to evaluate the velocity moments.
We assume that all the pseudo-fluid moments scale comparably,
\[
    n^*\sim \upar^* \sim \Tpar^* \sim \Tperp^* \sim q_\parallel^* \sim q_\perp^* 
    \sim P_\parallel^{\parallel*} \sim P_\parallel^{\perp*} \sim P_\perp^{\perp*} \sim \mathcal{O}(1).
\]
We now substitute these expansions into the GM hierarchy, Eq.~\eqref{eq:moment_hierarchy}, considering contributions up to order $\order{\tau^2}$.
This yields a system of equations composed of the density equation, $(p,j)=(0,0)$,
\begin{align}
    \ddt n^*
     + \pb{\phi}{n^*}  
     + \tau\,\pb{\lperp \,\phi}{\,\Tperp^*-n^*}
     + \sqrt{\tau}\,\bigl(\Cpar -  C_\parallel^B\bigr)\,\upar^*
    \nonumber\\[6pt]
     + \tau\,\Cper\,\bigl(\sqrt{2}\,\Tpar^* 
     + 2\,n^* 
     - \Tperp^*\bigr) 
     + \bigl(2\,\Cper  + R_N\,i\,k_y\bigr)\,\phi 
    \nonumber\\[-3pt]
     - \tau\,\Bigl(3\,\Cper  + \bigl[R_N+R_T\bigr]\,i\,k_y\Bigr)\,\lperp\,\phi 
     = \mathcal C_i^{00}  + \order{\tau^2},
    \label{eq:ni_eq_o2}
\end{align}
the parallel velocity equation, $(p,j)=(1,0)$,
\begin{align}
    \ddt \upar^*
     + \pb{\phi}{\upar^*} 
     + \tau\,\pb{\lperp\,\phi}{\,q_\perp^* - \upar^*} 
     + \sqrt{\tau}\,\Bigl(\bigl[\Cpar -  C_\parallel^B\bigr]\sqrt{2}\,\Tpar^*
     + \Cpar\,n^* 
    \nonumber\\[-3pt]
     - C_\parallel^B\,\Tperp^*\Bigr)
     + \tfrac{1}{\sqrt\tau}\,\Cpar\,\phi 
     - \sqrt\tau\,(\Cpar+C_\parallel^B)\,\lperp\,\phi 
     + \tau\,\Cper\,\bigl(\sqrt6\,q_\parallel^* + 4\,\upar^* - q_\perp^*\bigr)
    \nonumber\\[-3pt]
    = \mathcal C_i^{10}  + \order{\tau^{2}},
    \label{eq:upar_eq_o2}
\end{align}
the parallel temperature equation, $(p,j)=(2,0)$,
\begin{align}
    \ddt \Tpar^*
     + \pb{\phi}{\Tpar^*}
     + \tau\,\pb{\lperp\,\phi}{\,P_\parallel^{\perp*}-\Tpar^*} 
     + \sqrt{\tau}\,\Bigl(\sqrt{3}\,\bigl[\Cpar - C_\parallel^B\bigr]\,q_\parallel^*
     + \sqrt2\,\Cpar\,\upar^*\Bigr)
    \nonumber\\[-3pt]
     + \tau\,\Cper\,\Bigl(\sqrt{12}\,P_\parallel^{\parallel*}+6\,\Tpar^*
         + \sqrt{2}\,n^*\,-\,P_{\parallel}^{\perp*}\Bigr)
     + \Bigl(1-\tau\,\lperp\Bigr)\Bigl(\sqrt2\,\Cper  + \tfrac{\sqrt2}{2}\,R_T\,i\,k_y\Bigr)\,\phi 
    \nonumber\\[-3pt]
    = \mathcal C_i^{20}  + \order{\tau^2},
    \label{eq:Tpar_eq_o2}
\end{align}
and the perpendicular temperature equation, $(p,j)=(0,1)$,
\begin{align}
    \ddt \Tperp^*
     + \pb{\phi}{\Tperp^*}
     + \tau\,\pb{\lperp\,\phi}{\,n^* - 2\,\Tperp^* + 2\,P_\perp^{\perp*}}
     + \sqrt{\tau}\,\Cpar\,q_\perp^*
    \nonumber\\[-3pt]
     + \sqrt{\tau}\,C_\parallel^B\,\upar^*
     + \tau\,\Cper\,\bigl(\sqrt2\,P_\parallel^{\perp*}  + 4\,\Tperp^* 
      - n^* - 2\,P_\perp^{\perp*}\bigr)  
    \nonumber\\[-3pt]
     - \bigl(\Cper + R_T\,i\,k_y\bigr)\,\phi 
     + \tau\,\Bigl(5\,\Cper + \bigl[R_T+3\,R_N\bigr]\,i\,k_y\Bigr)\,\lperp\,\phi 
     = \mathcal C_i^{01}  + \order{\tau^2}.
    \label{eq:Tperp_eq_o2}
\end{align}
In addition, assuming adiabatic electrons, the GK quasi neutrality equation reduces to
\begin{equation}
    \Bigl(1 - 2\,\bigl[\lperp - \tau\,\lperp^2\bigr]\Bigr)\,\phi 
     - \langle \phi \rangle_{y z}  
     =  n^* + \tau\,\lperp\,\bigl(\Tperp^* - n^*\bigr) 
     + \order{\tau^2}.
    \label{eq:Poisson_eq_o2}
\end{equation}

Equations \eqref{eq:ni_eq_o2}--\eqref{eq:Tperp_eq_o2}, together with Eq.~\eqref{eq:Poisson_eq_o2}, contain higher-order GMs (such as $q_\parallel^*$, $q_\perp^*$, etc.), thus requiring additional assumptions for closure. \modifi{To close the system, we adopt a mixed-order truncation strategy: we retain $\order{\tau}$ terms in the density equation while dropping $\order{\tau}$ terms in the higher-order moment equations (parallel velocity, parallel and perpendicular temperatures).}
Specifically, by dropping $\order{\tau}$ terms in the parallel velocity, parallel temperature, and perpendicular temperature equations, we obtain,
\begin{align}
    & \ddt \upar^* + \pb{\phi}{\upar^*}
       + \sqrt{\tau}\,\Bigl(\bigl[\Cpar - C_\parallel^B\bigr]\sqrt2\,\Tpar^*
       + \Cpar\,n^*
       - C_\parallel^B\,\Tperp^*\Bigr)\nonumber\\
    &\qquad  + \frac{1}{\sqrt\tau}\,\Cpar\,\phi
        - \sqrt\tau\,(\Cpar + C_\parallel^B)\,\lperp\,\phi
        = \mathcal{C}_i^{10}  + \order{\tau},
    \label{eq:upar_eq_o1}\\[6pt]
    & \ddt \Tpar^* + \pb{\phi}{\Tpar^*}
       + \bigl(\sqrt2\,\Cper 
         + \tfrac{\sqrt2}{2}\,R_T\,i\,k_y \bigr)\,\phi
        = \mathcal C_i^{20} + \order{\tau^{1/2}},
    \label{eq:Tpar_eq_o1}\\[6pt]
    & \ddt \Tperp^* + \pb{\phi}{\Tperp^*}
        - \bigl(\Cper + R_T\,i\,k_y\bigr)\,\phi
        = \mathcal C_i^{01} + \order{\tau^{1/2}}.
    \label{eq:Tperp_eq_o1}
\end{align}

To account for collisions, we consider the gyro-averaged Dougherty operator projected over the Hermite--Laguerre basis \citep{Dougherty1964,Frei2022LocalMode}.
In the HEL, the collision operator terms \modifi{are given by}
\begin{align}
   \mathcal C_i^{00}  = & -\,\nu\,\tfrac{2}{3}\,\tau\,\lperp\,\Bigl(\sqrt2\,\Tpar^*  + \Tperp^*  + 5\,\lperp\,\phi \Bigr) 
    + \order{\tau^2}, 
   \\
   \mathcal C_i^{10}  = & \nu\,\order{\tau}, 
   \\
   \mathcal C_i^{20}  = & -\,\nu\,\tfrac{2}{3}\,\Bigl(2\,\Tpar^*
                  + \sqrt2\,\Tperp^*
                  + 2\sqrt2\,\lperp\,\phi\Bigr)
    + \order{\tau},
   \\
   \mathcal C_i^{01}  = & -\,\nu\,\tfrac{2}{3}\,\Bigl(\sqrt2\,\Tpar^*
                  + \Tperp^*
                  + 2\,\lperp\,\phi\Bigr)
    + \order{\tau},
\end{align}
where $\nu$ is the normalized ion--ion collision frequency.
Note that the Landau-based collision operator used in \cite{Ivanov2020ZonallyTurbulence} differs from the Dougherty model used here.
As a consequence, one should expect slight differences in the small-$\tau$ limit.
In the following, we refer to equations \eqref{eq:ni_eq_o2}, \eqref{eq:upar_eq_o1}--\eqref{eq:Tperp_eq_o1}, along with the quasi neutrality condition \eqref{eq:Poisson_eq_o2}, as the \emph{HEL--GM model}.

\modifi{The mixed-order closure employed in the HEL--GM model is an ad hoc truncation strategy designed to demonstrate that the Ivanov model emerges naturally from the GM hierarchy when appropriate limits are taken. This approach allows us to establish that the Ivanov fluid model is embedded within the more general GM hierarchy, thereby demonstrating that appropriate closure choices can recover existing reduced models while retaining the flexibility to extend beyond their original scope of validity.}

\subsection{Analytical equivalence with the Ivanov model in Z-pinch geometry}
\label{sec:Equiv_Zpinch}
We now illustrate that the HEL--GM model recovers the \emph{Ivanov model} when considering the Z-pinch magnetic geometry ($R_N=0$, $\Cper=-\,i\,k_y$, $\Cpar=1$, $C_\parallel^B=0$). 
Since the Ivanov model does not express the gyroaveraging operator in terms of Bessel functions, we directly expand the gyroaveraged distribution function $g_i(\bm R)$ (in gyrocenter coordinates $\bm R$) for small $\tau$ in terms of the distribution function in particle coordinates $f_i(\bm x)$,
\begin{align}
    g_i(\bm R) 
    &= \Bigl\langle\,f_i(\bm x)  - \bm\rho\cdot\nabla\,f_i(\bm x) 
        + \tfrac12\,\bm\rho\,\bm\rho : \nabla\nabla\,f_i(\bm x)\Bigr\rangle 
     + \order{\tau^2},\nonumber\\
    &= g_i(\bm x)  + \tfrac{\tau}{2}\,\wperp\,\nabla_\perp^2\,g_i(\bm x) 
     + \order{\tau^2},
\end{align}
where $\bm \rho =  \sqrt{2\tau}\wperp\hat{\bm b}$ is the gyrocenter displacement vector, with $\hat{\bm b}$ the unit vector along the magnetic field, and $\nabla_\perp^2$ the perpendicular Laplacian operator.
Recalling that the gyro-averaging operator, $\langle\,\cdot\,\rangle$, satisfies $\langle\bm\rho\rangle=0$ and $\langle\bm\rho\,\bm\rho\rangle = \tau\,\wperp\,\mathbf{I}_\perp$, with $\mathbf{I}_\perp$ the perpendicular projection operator, a pseudo-fluid moment, e.g., $n^*$, can be expressed in the particle coordinate system via
\begin{align}
   n^*(\bm R)  &=  \iint \Bigl[g_i(\bm x) 
        + \tfrac{\tau}{2}\,\wperp\,\nabla_\perp^2\,g_i(\bm x)\Bigr] 
       \,\mathrm d\wperp\,\mathrm d\spar
        + \order{\tau^2}
    \nonumber\\
    &=n^*(\bm x) 
         + \tfrac{\tau}{2}\,\nabla_\perp^2\bigl(n^* - T_\perp^*\bigr)(\bm x) 
         + \order{\tau^2},
\end{align}
where we assume commutation between the velocity-space integration and the perpendicular Laplacian operator. This assumption is valid in the local approximation, where the perpendicular gradients are considered constant over a Larmor radius.
The gyrocenter-to-particle coordinate transformation does not affect the higher-order GMs, as the HEL-GM scaling neglects the $\order{\tau}$ terms in the parallel velocity, parallel temperature, and perpendicular temperature equations.

The Ivanov model is then obtained by rewriting the HEL-GM model, Eqs.~\eqref{eq:ni_eq_o2} and \eqref{eq:upar_eq_o1}--\eqref{eq:Tperp_eq_o1}, for the following fluid moments,
\begin{equation}
    n(\bm x) = n^*(\bm x), 
    \quad
    u_\parallel(\bm x) = \frac{\sqrt2}{2}\,\upar^*(\bm x), 
    \quad
    T(\bm x) = \frac{1}{2} \left[\sqrt{2} T_{\parallel}^*(\bm x) - T_{\perp}^*(\bm x)\right] - n^*(\bm x),
\end{equation}
and considering a Z-pinch geometry $(\Cper=-\,i\,k_y,\,\Cpar=1,\,C_\parallel^B=0)$ and $\lperp=\nabla_\perp^2/2$. 
\modifi{To obtain analytical equivalence with the Ivanov model, it is necessary to rescale the variables with the asymptotically small parameter $\tau$. Consequently, this procedure modifies the orderings between the evolved moments, which is justified considering the high-collisionality, long-wavelength, and large aspect ratio limits used in the Ivanov model.}
Specifically, the rescaling is defined as $\hat z = 2z$, $\hat \phi=\tau \phi/2$, $\hat u_\parallel=u_\parallel/\tau$, $\hat T = \tau T/2$, and $\kappa_T=\tau R_T/2$.
Finally, the collisionality parameter of the Ivanov model is linked to the HEL-GM collision frequency parameter $\nu$ using the relation
\begin{equation}
    \chi = c_f\frac{2}{3}\tau\nu,
\end{equation}
where we introduce an empirical factor $c_f$ to account for the differences between the collision models.
This empirical value is determined by a direct comparison of linear growth rates and nonlinear saturation levels between our HEL--GM system and the results reported in \cite{Ivanov2022DimitsTurbulence}, ensuring quantitative agreement across the relevant parameter space. We set $c_f=4$ for the rest of this work (see Fig.~\ref{ch6_fig:heat_flux_2D}).

In summary, in the present work, we consider three models: (i) the GK model, Eqs.~\eqref{eq:moment_hierarchy} and \eqref{eq:poisson_moments_adiabe}, solved using the GM approach and a Dougherty collision model, which provides the most complete description of the plasma dynamics considered here; (ii) the HEL--GM model, consisting of Eqs.~\eqref{eq:ni_eq_o2}, \eqref{eq:upar_eq_o1}--\eqref{eq:Tperp_eq_o1}, and \eqref{eq:Poisson_eq_o2}, which is the GM hierarchy closed by the hot-electron limit using a mixed-order closure (i.e., $\mathcal{O}(\tau)$ terms are retained only in the density equation, while higher-moment equations are truncated at a reduced order for consistency); and (iii) the Ivanov fluid model in a Z-pinch geometry (see Eqs. (2.4-2.6) in \cite{Ivanov2022DimitsTurbulence}), which considers HEL moments of the Landau collision operator.
In Sec. \ref{sec:linear_results}, we show that the \gyacomo code can effectively retrieve the HEL-GM model when considering a sufficiently small temperature ratio $\tau$ and when the gradient and collisionality parameters are scaled accordingly. These simulations are referred to as hot electron limit \gyacomo simulations (HEGS), \modifi{as they do not solve the HEL-GM system directly but asymptotically from the GM equations \eqref{eq:moment_hierarchy}.}

\section{Verification of the HEL-GM closure}
\label{sec:linear_results}

\modifi{The goal of this section is to demonstrate that a GK code, such as \gyacomo, can recover Ivanov's reduced fluid model when considering the appropriate parameter regime. We highlight this point through three main verification steps.}
First, we evaluate the growth rates of the instabilities present in the two-dimensional Z-pinch geometry with the \gyacomo code, varying the temperature ratio and the number of evolved GMs to show proper convergence to the HEL-GM limit.
\modifi{Second, we compare \gyacomo results with the linear results of \cite{Ivanov2020ZonallyTurbulence} and \cite{Ivanov2022DimitsTurbulence} to verify that the same closed set of equations is obtained when considering $\tau\ll1$.
Third, we benchmark our three dimensional simulations against the results of \cite{Ivanov2022DimitsTurbulence}.}
In addition, we assess the impact of the Dougherty collision operator on the linear growth rates.

We start by focusing on the linear predictions. The HEL-GM system exhibits several instabilities in the Z-pinch geometry, including the slab ITG (sITG) and curvature-driven ITG (cITG) modes \citep{rudakov1961slabITG,Pogutse1968curv_ITG}. On the other hand, the entropy mode \citep{Ricci2006,Kobayashi2015a,Hoffmann2023GyrokineticOperators} is not present due to the adiabatic electron assumption.
The cITG mode arises from the presence of curvature or a perpendicular gradient of the local magnetic field, developing primarily in the poloidal direction with negligible parallel dependence.
When $k_\parallel \neq 0$, sITG modes emerge due to the coupling between density, parallel velocity, and temperature fluctuations, propagating predominantly in the parallel direction.

\begin{figure}
    \centering 
    \includegraphics[width=\linewidth]{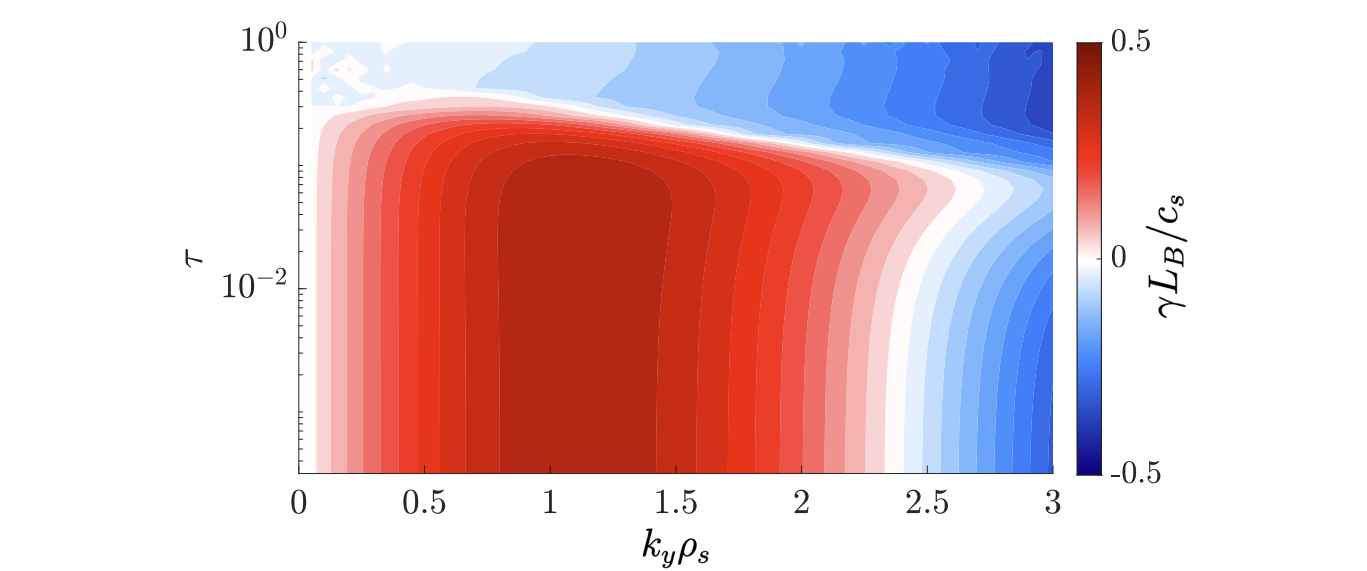}
    \caption{2D Z-pinch ITG linear growth rates with respect to the $\tau$ parameter obtained with \gyacomo for $d_{\max}=2$, $\kappa_T=0.36$ and $\chi=0.1$.}
    \label{ch6_fig:linear_convergence_tau}
\end{figure}

\begin{figure}
    \centering 
    \includegraphics[width=0.7\linewidth]{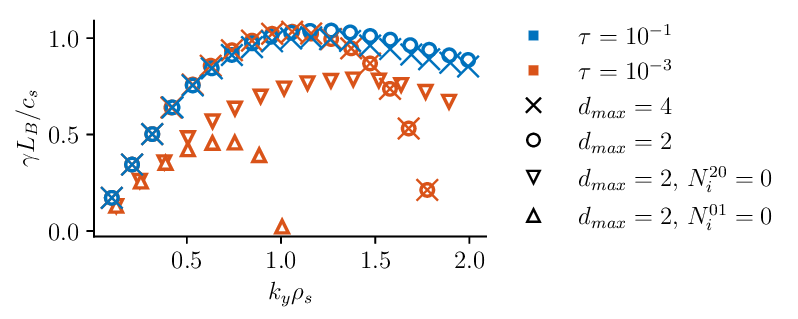}
    \caption{ITG growth rates obtained with \gyacomo in two-dimensional Z-pinch geometry for 
    $d_{\max}=4$ (crosses),
    $d_{\max}=2$ (circles), 
    $d_{\max}=2$ without $N_i^{20}$ (down triangles), 
    and 
    $d_{\max}=2$ without $N_i^{01}$ (up triangles), 
    using $\tau=10^{-1}$ (blue) and $\tau=10^{-3}$ (red). The gradient and collision are set to $\kappa_T=1.0$, and $\chi=0$, respectively.}
    \label{ch6_fig:linear_HEL}
\end{figure}

We evaluate the linear growth rate of the Z-pinch instabilities using the \gyacomo code as a function of $\tau$ and for different GM sets. The \gyacomo temperature gradient parameter is scaled accordingly: $R_T=\tau \kappa_T$, where $\kappa_T$ is Ivanov's temperature gradient parameter.
Figure \ref{ch6_fig:linear_convergence_tau} illustrates the dependence of the ITG linear growth rates on $\tau$, setting $d_{\max}=2$ (4 GMs). We observe that the growth rates become independent of $\tau$ for $\tau\lesssim 10^{-2}$, despite an increasing temperature gradient $R_T$, which indicates that the system is reaching the HEL-GM limit.
Furthermore, Fig. \ref{ch6_fig:linear_HEL} shows that identical results are obtained when the number of GMs is increased from 4 GMs ($d_{\max}=2$) to 9 GMs ($d_{\max}=4$), when considering a sufficiently small value of $\tau$.
We also report that considering a smaller GM set, i.e., fewer than 4 GMs ($d_{\max}<2$), does not reproduce the same growth rates, highlighting the importance of retaining the $N_i^{20}$ and $N_i^{01}$ GMs in the HEL-GM closure.
These points demonstrate that the $N_i^{00}$, $N_i^{10}$, $N_i^{20}$, and $N_i^{01}$ GM system is a closed set of equations when $\tau$ is sufficiently small and when the temperature gradient is scaled accordingly.
Following this analysis, we choose $\tau=10^{-3}$ and $d_{\max}=2$ in the HEGS presented hereafter.

\begin{figure}
    \centering 
    \includegraphics[width=0.7\linewidth]{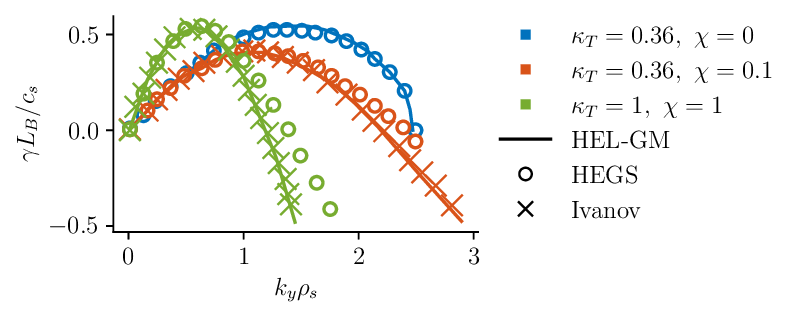}
    \caption{ITG growth rates vs. poloidal wave number in two-dimensional Z-pinch geometry obtained with the HEL-GM linear solver (solid lines), \gyacomo with $d_{\max}=2$ and $\tau=10^{-3}$ (circles), and by \cite{Ivanov2020ZonallyTurbulence} (stars) for three parameter sets: $\kappa_T=1$, $\chi=1$ (green); $\kappa_T=0.36$, $\chi=0.1$ (red); $\kappa_T=0.36$, $\chi=0$ (blue).}
    \label{ch6_fig:linear_ivanov_conv}
\end{figure}

We now compare the growth rates obtained by HEGS with those from \cite{Ivanov2020ZonallyTurbulence} in Fig. \ref{ch6_fig:linear_ivanov_conv}.
Good agreement is observed, particularly at lower collisionalities, suggesting that the collision operator is the primary source of discrepancy between the two models.
To further examine the impact of using the Dougherty operator, which retains higher order $\tau$ terms in \gyacomo, we solve the eigenvalue problem associated with the HEL-GM linear system, which contains an $\order{\tau}$ Dougherty model.
The eigenvalues of the HEL-GM model exhibit closer agreement with \cite{Ivanov2020ZonallyTurbulence} than the HEGS (see Figure \ref{ch6_fig:linear_ivanov_conv}) for finite collisionalities, suggesting that the differences arise primarily from the HEGS collision model.
When considering a collisionless case, not explored in \cite{Ivanov2020ZonallyTurbulence}, we find perfect agreement between the HEL-GM solver and the HEGS, confirming that the higher-order terms of the Dougherty operator are indeed the source of the observed discrepancies.


\begin{figure}
    \centering 
    \includegraphics[width=\linewidth]{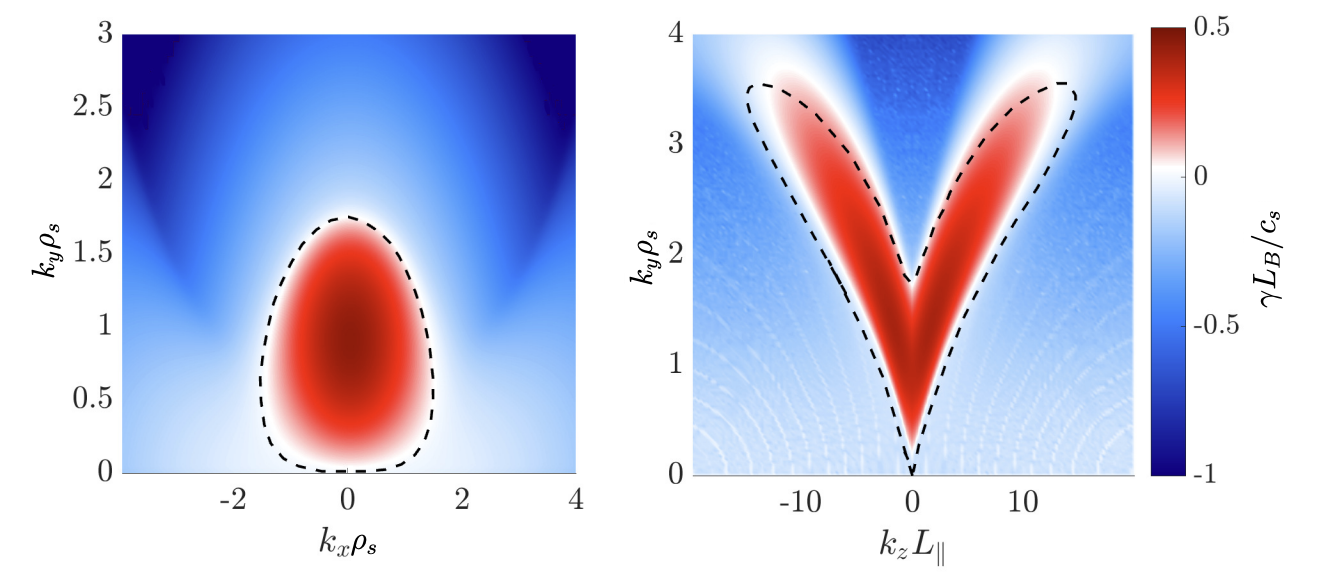}
    \caption{ITG linear growth rates in the 3D Z-pinch geometry for $\kappa_T=1$ and $\chi=0.1$ obtained with HEGS (setting \PJ{2}{1} and $\tau=10^{-3}$ in \gyacomo), and comparison with the stability limit obtained in \cite{Ivanov2022DimitsTurbulence} (dashed line).}
    \label{ch6_fig:3DZP_lin_growthrate}
\end{figure}

Finally, we explore the linear sITG and cITG instabilities in Fig. \ref{ch6_fig:3DZP_lin_growthrate} by examining the growth rates for different radial and parallel mode numbers, considering $\kappa_T=1$ and $\chi=0.1$, and introducing $L_\parallel=2L_z$ because of the different normalization considered by the HEL-GM and the Ivanov models.
\modifi{The results show close agreement with those of \cite{Ivanov2022DimitsTurbulence}, with discrepancies observed at large $k_y$, which are similar to those seen in the two-dimensional case, and at large $k_z$, which may be attributed to the finite difference scheme used in \gyacomo in the parallel direction.}

We now turn to nonlinear simulations. We first consider two-dimensional simulations on a domain of size $L_x=100$ and $L_y=150$, with a resolution of $N_x=N_y=256$.
We set the temperature gradient values to $\kappa_T=0.36$, $1$, and $2$, and collision frequencies between $\chi=10^{-3}$ and $10^1$. 
The hyperdiffusion parameter is set to \modifi{$\mu_{hd}=1.0$, ensuring, for each simulation, that the linear growth rates of the cITG instability are not affected by the hyperdiffusion.}

\begin{figure}
    \centering
    \includegraphics[width=\linewidth]{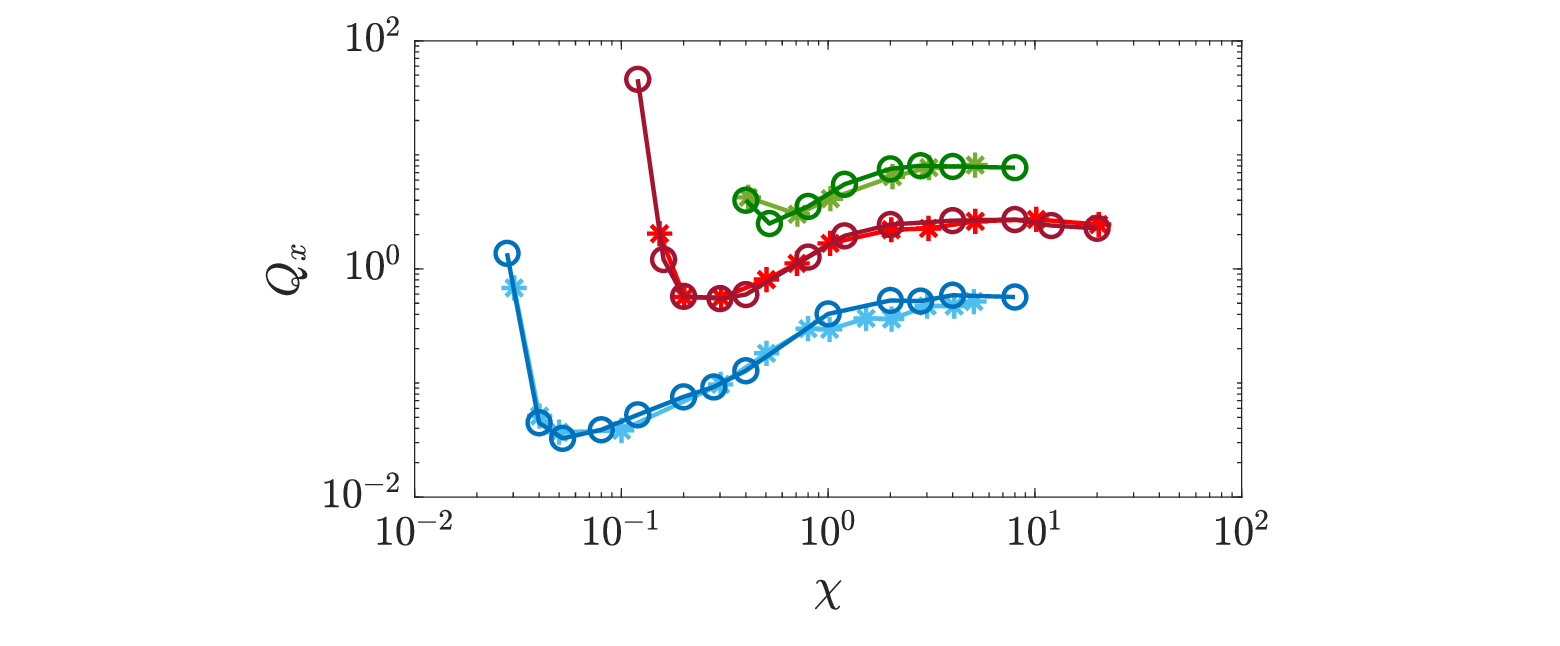}
    \caption{Saturated heat flux obtained with HEGS (circles) and from \cite{Ivanov2020ZonallyTurbulence} (stars) for $\kappa_T=0.36$ (blue), $\kappa_T=1$ (red), and $\kappa_T=2$ (green), in the two-dimensional Z-pinch geometry.}
    \label{ch6_fig:heat_flux_2D}
\end{figure}

Figure \ref{ch6_fig:heat_flux_2D} compares the heat fluxes obtained in the HEGS with the results from \cite{Ivanov2022DimitsTurbulence}.
We report excellent quantitative agreement for all considered temperature gradients, confirming that the HEGS captures the same nonlinear physics as the Ivanov model.
In the high collisionality regime, the turbulent heat flux saturates to a value that increases with the temperature gradient, reflecting the linear growth rate dependence.
The heat flux is significantly reduced with the decrease of collisionality, up to a threshold value below which fully developed cITG turbulence fails to saturate, \modifi{due to the absence of three-dimensional effects} \citep{Barnes2011CriticallyPlasmas}.

We finally aim to verify if the HEGS can reproduce the results of \cite{Ivanov2022DimitsTurbulence} when simulating turbulence in a three dimensional Z-pinch geometry. 
We use the \gyacomo code, setting $L_x=L_y=80$ with a resolution of $N_x=N_y=128$.
We set the parallel resolution to $N_z=16 \lceil N_{pol}\rceil$ for $\kappa_T=0.36$, $N_z=50 \lceil N_{pol}\rceil$ for $\kappa_T=0.8$, and $N_z=100 \lceil N_{pol}\rceil$ for $\kappa_T=3.0$. Here, $\lceil N_{pol}\rceil$ denotes rounding $N_{pol}$ up to the nearest integer.
It is worth noting that the higher considered temperature gradient leads to unsaturated turbulence in the two-dimensional system (see Sec. \ref{sec:nonlinear_results}), which is mitigated by the excitation of sITG modes at a finite parallel wavenumber $k_z$ (see Fig. \ref{ch6_fig:3DZP_lin_growthrate}).

\begin{figure}
    \centering
    \includegraphics[width=1\linewidth]{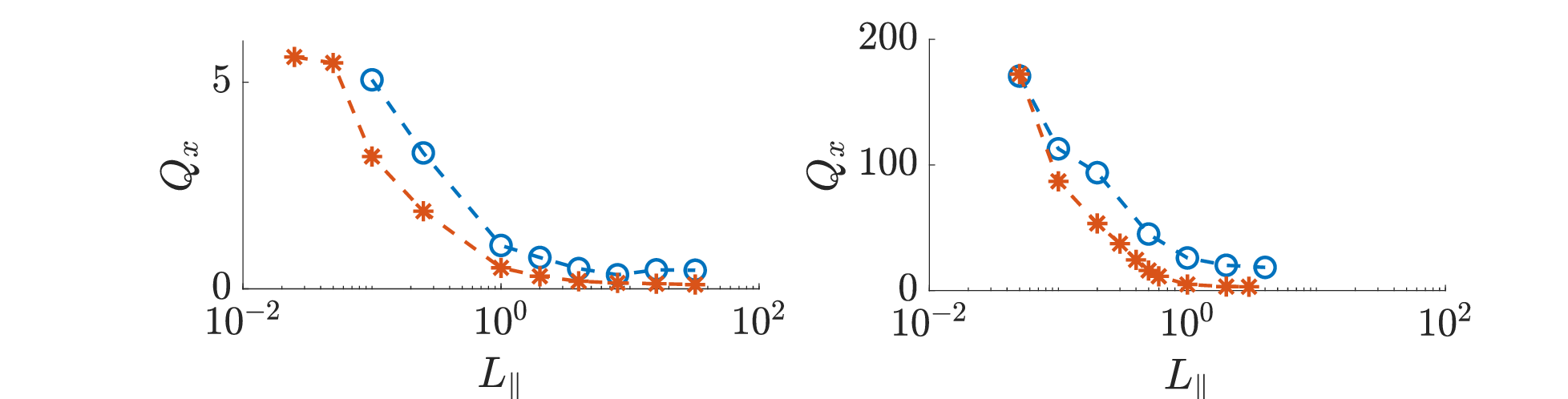}
    \caption{Saturated heat flux level with respect to the length of the flux tube domain in the parallel direction for $\kappa_T=0.8$ (left) and $\kappa_T=3.0$ (right), setting $\chi=0.1$.
    We compare the results from the HEGS (blue) and \cite{Ivanov2022DimitsTurbulence} (red).}
    \label{ch6_fig:Q_x_vs_Lpar}
\end{figure}

Figure \ref{ch6_fig:Q_x_vs_Lpar} shows that the HEGS predictions are quantitatively close to those of \cite{Ivanov2022DimitsTurbulence}, but with a slightly higher transport level for almost all $L_\parallel$ values considered.
This difference may stem from different tuning of numerical diffusion parameters, \modifi{discrepancies in the collision operators}, but also from the different representations of the parallel direction.
\gyacomo does not use a spectral representation of $z$, in contrast to the method used in \cite{Ivanov2022DimitsTurbulence}.
Despite this discrepancy, we observe a stabilization of the heat flux value around the same parallel length of the domain, indicating an agreement in capturing the main features of the turbulence dynamics.

\hyphenation{collisionality}

\section{Nonlinear transport physics in Z-pinch configurations}
\label{sec:nonlinear_results}

In this section, we analyze the HEGS nonlinear results, focusing on the Z-pinch geometry, first in two dimensions and then in three.

\begin{figure}
    \centering
    \includegraphics[width=\linewidth]{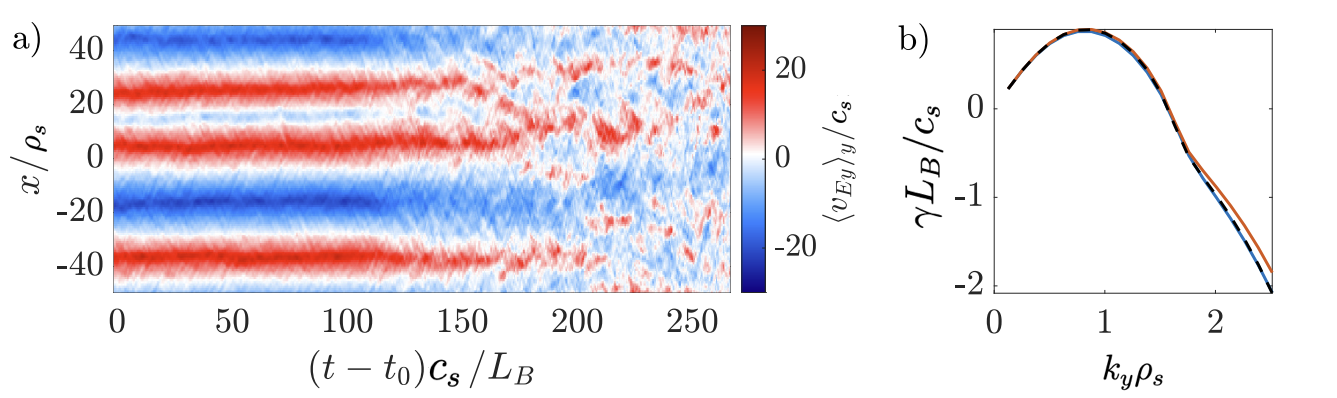}
    \caption{(a) ZFs velocity averaged along the binormal direction, $\langle v_{Ex}\rangle_y$, for $\kappa_T=1.2$, obtained by setting $\chi=0.16$ at a restart of a $\chi=0.2$ simulation.
    (b) Linear growth rates of the ITG instability for $\chi=0.20$ (blue) and $\chi=0.16$ (red), setting $\kappa_T=1.2$ and the hyperdiffusion $\mu_{hd}=1$, used in the nonlinear case.
    The black curve corresponds to the $\chi=0.16$ case with a $20\%$ increase in the hyperdiffusion parameter \modifi{$\mu_{hd}$}.}
    \label{ch6_fig:2D_blow_up}
\end{figure}

Two-dimensional simulations show that the level of transport increases with the strength of the temperature gradient and, most interestingly, blows up at low collisionality.
To understand the mechanisms that lead to a blow up, we consider a simulation in its steady state with parameters $\kappa_T=1.2$ and $\chi=0.2$. 
We then restart the simulation, introducing a $20\%$ reduction in the collisionality value. 
This leads to a destabilization of the zonal flows (ZFs) and a blow-up of the heat flux (see Fig. \ref{ch6_fig:2D_blow_up}a).
We note that this collisionality decrease barely affects the linear growth rates of the cITG instability, as shown in Fig. \ref{ch6_fig:2D_blow_up}b. (We confirm that the differences in the growth rate have a negligible effect by carrying out nonlinear simulations where the value of hyperdiffusion is increased so that the linear growth rate of the low collisionality case matches that of the higher collisionality simulation, while a blow up state is still observed.)
Similarly, we note that the HEL-GM eigenvalue solver reports a negligible effect of collisionality on the subdominant eigenvalues. However, a blow-up of the transport is observed when the collisionality value is reduced. 
Hence, we conclude that the blow-up is not due to a change in the linear properties of the driving instability but is rather the result of a change in the nonlinear saturation mechanism of the driving instability.

At a collisionality just above the blow-up threshold, the turbulence presents a bursty behavior, with intermittent phases of high and low transport reminiscent of a predator-prey cycle, where ZFs (the "predator") suppress turbulence (the "prey"), and weakened ZFs allow turbulence to grow again. This cyclical interaction is typical of ZF turbulence dynamics \citep{Kobayashi2015,Ivanov2020ZonallyTurbulence,Hoffmann2023GyrokineticOperators}.
On the other hand, at larger collisionality, the bursty behavior is replaced by a turbulence-dominated state where the ZF amplitude is significantly reduced in comparison to the fluctuation amplitude.

The sudden increase in heat flux when collisionality is below a threshold value is in agreement with the results of \cite{Ivanov2020ZonallyTurbulence}. 
At low collisionality, \cite{Ivanov2020ZonallyTurbulence} reports a negative turbulent viscosity value, which implies that turbulence no longer strengthens the ZFs, thus removing the saturation mechanism for the growth of the primary instability.
While the agreement between the HEGS and \cite{Ivanov2020ZonallyTurbulence} suggests that the HEGS captures the same physics, this mechanism may be limited to the HEL model, as it does not agree with more complete GK models.
\cite{Ricci2006a,Hallenbert2022,Hoffmann2023GyrokineticOperators} show a steady increase of transport with respect to increasing collisionality in two-dimensional Z-pinch GK simulations and do not report a blow-up state at low collisionality.
Additionally, \cite{Sarazin2021KeyPlasmas} demonstrate, by using GK simulations carried out with the \textsc{Gysela} code \citep{grandgirard2016}, that a transition to fully developed turbulence can be observed at low collisionality without a change of sign of the turbulent viscosity.

The three-dimensional geometry allows for the presence of modes with $k_\parallel \neq 0$, enabling turbulent eddies to lose correlation along the parallel direction.
This decorrelation reduces the parallel extension of an eddy and, as a consequence, its ability to transport energy radially.
On the other hand, the extension of the parallel length can increase the heat flux by destabilizing $k_\parallel\neq 0$ modes.
This is observed in \cite{Volcokas2023UltraTokamaks}, where the relationship between parallel domain length and heat flux is investigated by considering CBC GENE simulations at low magnetic shear.
When considering an adiabatic electron response, \cite{Volcokas2023UltraTokamaks} reports that the eddy correlation length in the parallel direction is strongly reduced.
In addition, a monotonic decrease of the heat flux is observed when the parallel elongation of the domain is extended.

The saturated transport level is reduced by increasing the parallel length until it reaches an asymptotic value.
This behavior, observed in Fig. \ref{ch6_fig:Q_x_vs_Lpar}, recalls the findings of \cite{Volcokas2023UltraTokamaks}, indicating that the HEGS captures the main features of the parallel decorrelation mechanism.
When the \modifi{parallel extension of the simulation domain is short}, turbulent eddies can \textit{self-interact}, i.e., interact with themselves through the periodic boundary conditions imposed along the parallel direction, allowing them to span the entire parallel extent of the domain. This leads to a higher transport level, closely resembling the two-dimensional limit where $k_\parallel=0$ is imposed.
As the parallel extension of the domain approaches the typical eddy correlation length, $L_\parallel \sim 32$, finite $k_\parallel$ fluctuations emerge.
Once the parallel dimension exceeds several correlation lengths, eddies are no longer able to self-interact, and their extension along $z$ saturates, as does the heat flux level.

\begin{figure}
    \centering
    \includegraphics[width=1\linewidth]{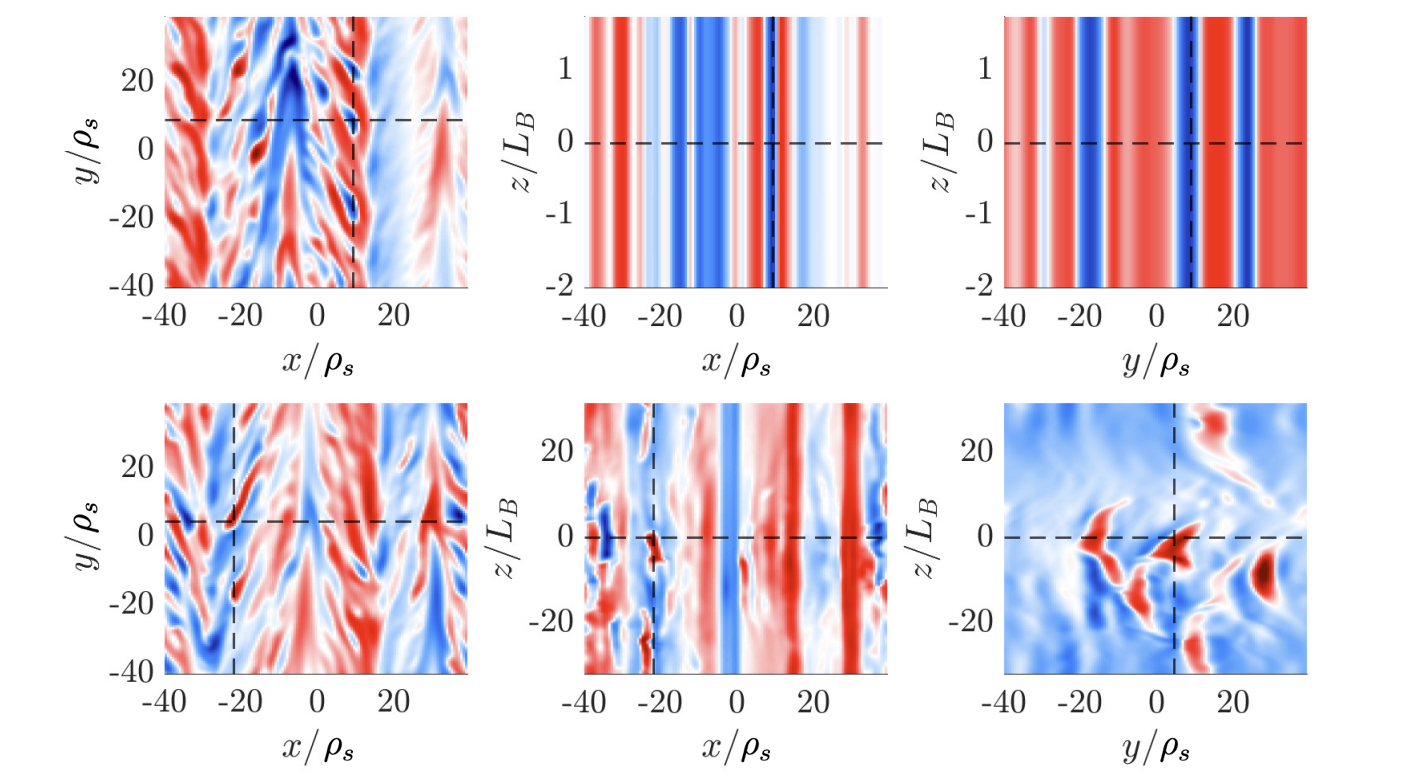}
    \caption{Snapshots of the temperature fluctuations during a burst of transport in Z-pinch ITG turbulence simulations for $L_\parallel = 4$ (top) and $L_\parallel = 128$ (bottom) obtained with \textsc{Gyacomo}, setting $\kappa_T=0.36$, $\chi=0.1$ and $\tau=10^{-3}$. 
    The dashed lines indicate the intersection between the three planes of the same row.}
    \label{ch6_fig:snapshots_3D_ferdinons}
\end{figure}

Figure \ref{ch6_fig:snapshots_3D_ferdinons} illustrates the decorrelation mechanism by comparing snapshots of the temperature fluctuations in simulations with domains of different extensions along the parallel direction in the weak turbulence regime ($\kappa_T=0.36$).
In the case of a short parallel length ($L_\parallel=8$), turbulent eddies extend along the entire domain, indicating strong correlation along the magnetic field line.
On the other hand, in a longer parallel domain ($L_\parallel=32$), the eddies lose phase coherence along $z$ and break into shorter, partially decorrelated structures. As soon as the domain exceeds a few parallel correlation lengths, further increases of $L_\parallel$ only weakly affect the time-averaged heat flux, which approaches its asymptotic value. In this regime, the dynamics transitions from isolated, domain-filling transport bursts to a superposition of smaller, spatially separated bursts and quiescent patches at different $z$ locations; their temporal dephasing smooths the global response while yielding a comparable average heat flux.

\begin{figure}
    \centering
    \includegraphics[width=1\linewidth]{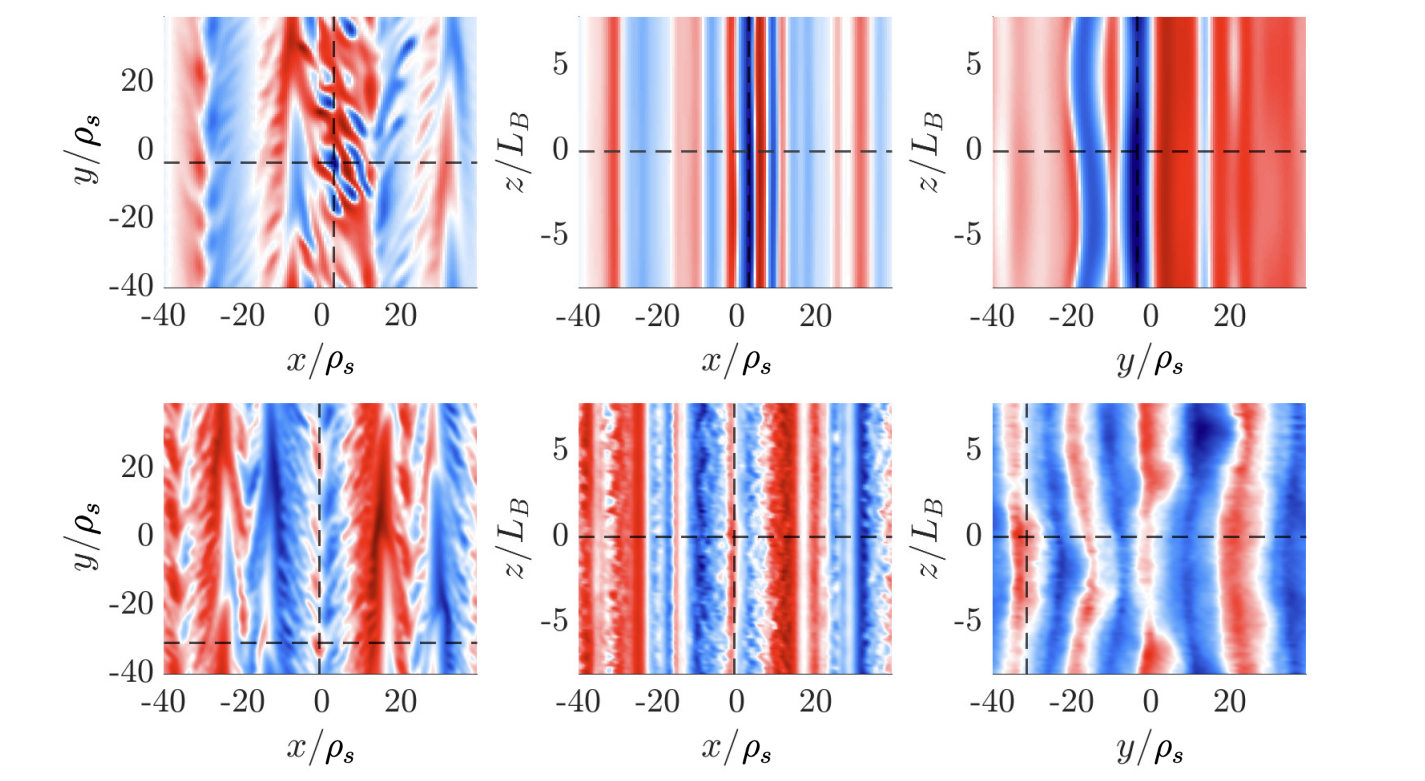}
    \caption{Snapshots of the temperature fluctuations during a burst of transport in Z-pinch ITG turbulence simulations for $\kappa_T=0.36$ (top) and $\kappa=0.8$ (bottom) obtained with \textsc{Gyacomo}, setting $L_\parallel=32$, $\chi=0.1$ and $\tau=10^{-3}$.
    The dashed lines indicate the intersection between the three planes of the same row.}
   \label{ch6_fig:3D_snapshot_kt_0.8_chi0.1}
\end{figure}

\begin{figure}
    \centering
    \includegraphics[width=1\linewidth]{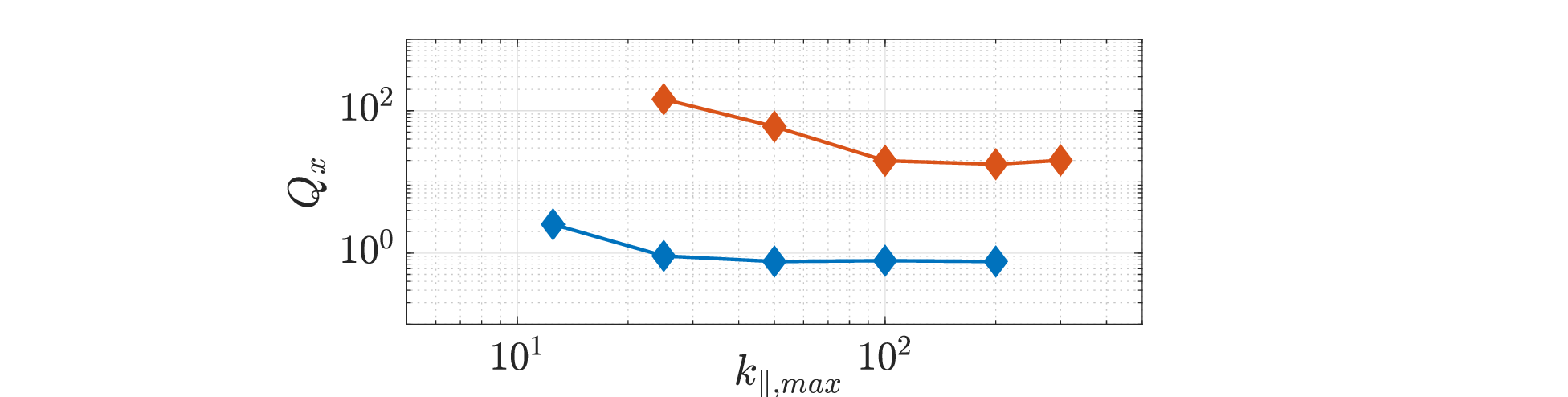}
    \caption{Convergence of the heat flux with respect to the parallel resolution in the three-dimensional Z-pinch geometry for $\kappa_T=0.8$ (blue) and $\kappa_T=3.0$ (red), setting $\chi=0.1$ and $L_\parallel=2$.}
    \label{ch6_fig:3D_HF_kpar_conv}
\end{figure}

Finally, we compare the simulations that display weak and strong turbulence (specifically, $\kappa_T=0.38$ and $\kappa_T=0.8$). 
Fig. \ref{ch6_fig:3D_snapshot_kt_0.8_chi0.1} presents snapshots of the turbulent temperature fluctuations for these two cases, setting $\chi=0.1$ and $L_\parallel=32$. 
Small parallel scale turbulence develops along the parallel direction for $\kappa_T = 0.8$ as a result of the excitation of sITG modes, in contrast to the weak turbulence regime.
These modes are responsible for the decorrelation of the turbulent eddies along the magnetic field line, reducing their parallel correlation length significantly compared to the weakly turbulent case.
These sITG modes are marginal in the weakly turbulent regime, where two-dimensional cITG modes with high parallel correlation dominate the dynamics.
We note that the saturated turbulent heat flux level is highly sensitive to the parallel resolution, highlighting the importance of accurately resolving the small parallel scales associated with the sITG modes (see Fig. \ref{ch6_fig:3D_HF_kpar_conv}).
When considering a larger temperature gradient, a larger parallel resolution is required to reach a saturated state, as the maximal unstable parallel mode number increases (see Fig. \ref{ch6_fig:3DZP_lin_growthrate}).

\section{Tokamak geometry and finite temperature ratio}
\label{sec:HEL_CBC}
\modifi{We now investigate the accuracy of the HEL closure in a tokamak geometry.} We perform simulations using \gyacomo in two different ways: (i) with the parameters identified in the previous sections to reach the Ivanov fluid model using the HEL, i.e., the HEGS, and (ii) with \gyacomo GK simulations evolving higher order moments and setting $\tau=1$. The GK simulations are performed setting $d_{\max}=4$, as \cite{Hoffmann2023GyrokineticShift} show that this is sufficient for the numerical convergence of the results. The resolution of the simulations is $(N_x,N_y,N_z)=(128,64,24)$ with a domain size of $L_x=L_y=120$ and $L_z=2\pi$.

We consider the parameters of the CBC, a standard test case for GK codes \citep{Lin1999EffectsTransport,Dimits2000ComparisonsSimulations}, using the tokamak $s-\alpha$ geometry with a safety factor $q_0=1.4$, a local magnetic shear $\hat s = 0.8$, and an inverse aspect ratio $\epsilon=0.18$.
The ion temperature gradient is set to $\kappa_T = 3.5$, which corresponds to $R_T=7$ for $\tau=1$, and a finite collision parameter $\chi = 0.02$ is used to facilitate the convergence of the GM hierarchy \citep{Hoffmann2023GyrokineticOperators}.
These parameters are based on a DIII-D tokamak discharge in the core plasma region \citep{Greenfield1997EnhancedDIII-D}, where the electron-to-ion temperature ratio is typically of order $\tau\sim1$, which does not satisfy the HEL assumption.
We explore the HEL as an alternative to the truncation closure scheme, which has limited accuracy when considering a small number of GMs \citep{Hoffmann2023GyrokineticShift}.

\begin{figure}
    \centering
    \includegraphics[width=0.5\linewidth]{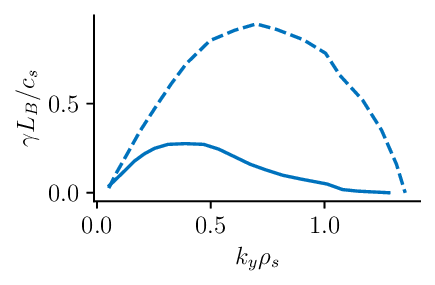}
    \caption{\modifi{Linear growth rates of ITG simulations in the $s-\alpha$ geometry with the GK simulations (solid) and HEGS (dashed).}}
    \label{ch6_fig:lin_CBC}
\end{figure}
\begin{figure}
    \centering
    \includegraphics[width=0.8\linewidth]{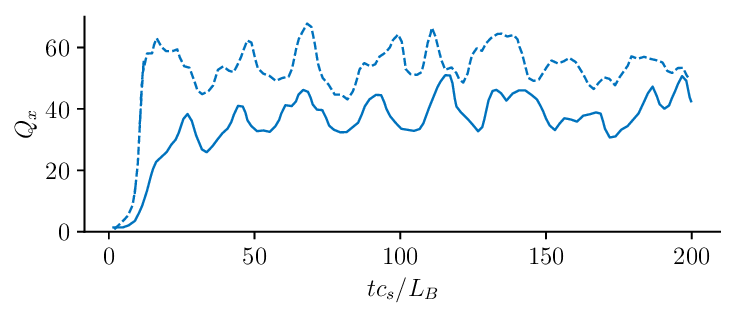}
    \caption{\modifi{Time traces of the radial heat flux of the Cyclone Base Case with the GK simulations (solid) and HEGS (dashed).}}
    \label{ch6_fig:HF_time_traces_CBC}
\end{figure}

CBC turbulence is ITG-driven and exhibits a Dimits shift when $\kappa_T$ is reduced \citep{Dimits2000ComparisonsSimulations}. The linear results (Fig.~\ref{ch6_fig:lin_CBC}) show that the HEGS growth rates are consistently higher than those of the GK simulations, peaking at $(k_y\rho_s,\gamma L_B/c_{s})\approx(0.75,0.9)$ versus $(0.5,0.25)$. The HEGS also sustains a broader unstable spectrum, similar to the trends seen with a low $d_{\max}$ truncation \citep{Hoffmann2023GyrokineticShift}. The nonlinear heat-flux time traces (Fig.~\ref{ch6_fig:HF_time_traces_CBC}) likewise show higher transport for the HEGS, but the relative increase is smaller than the linear growth-rate discrepancies, suggesting a limited sensitivity of the saturated flux to the additional small-scale linear drive. The temporal correlations are comparable, indicating qualitatively similar turbulence dynamics.

\begin{figure}
    \centering
    \includegraphics[width=1\linewidth]{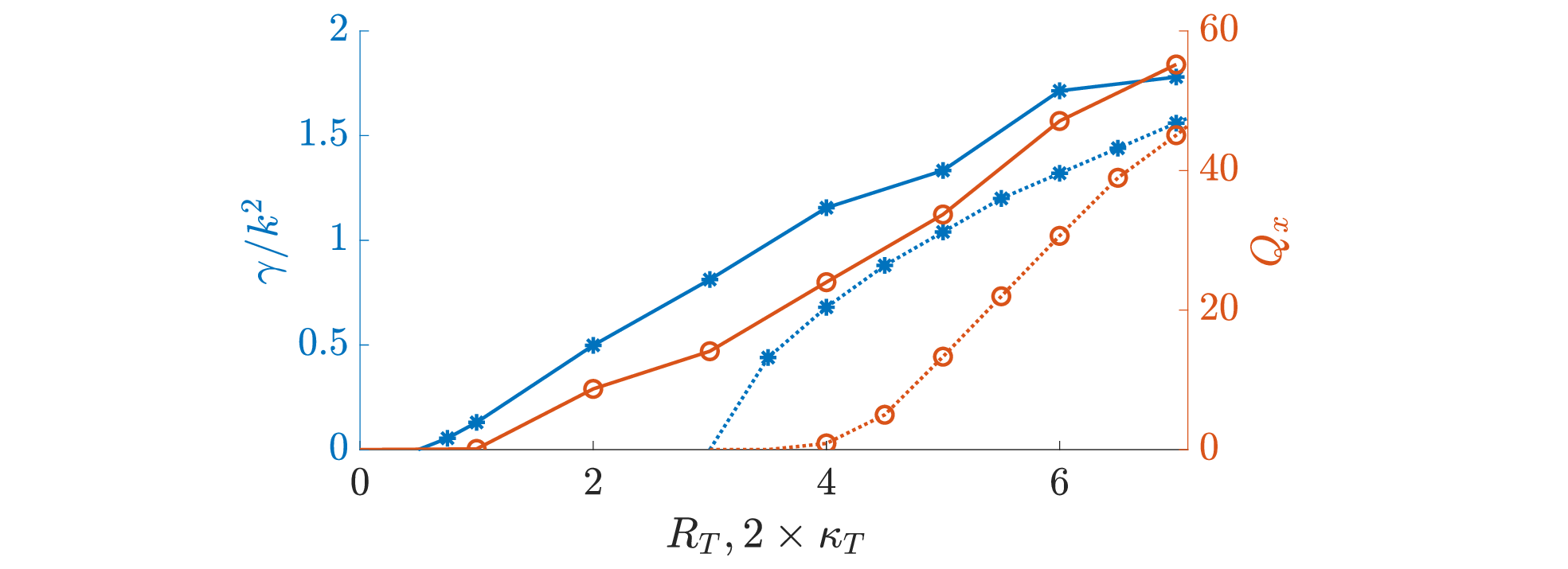}
    \caption{Mixing length estimate $\gamma/k^2$ of the maximal growth rate (blue) and the saturated heat flux level (red) for different temperature gradient values. 
    We compare the HEGS (solid) \modifi{with the GK simulations} (dashed).}
    \label{ch6_fig:dimits_shift_comparison_CBC_one_fig}
\end{figure}

We now investigate the capability of the HEGS in predicting a Dimits shift, comparing it to the GK simulations in Fig. \ref{ch6_fig:dimits_shift_comparison_CBC_one_fig}.
We evaluate the mixing length estimate $\gamma/k^2$ of the maximal growth rate and compare it to the saturated heat flux level for different temperature gradient values.
While the GK simulations exhibit a clear Dimits shift, the HEGS does not, as a non-vanishing heat flux level is observed very close to the linear threshold. \modifi{Instead, bursts of transport are observed with a periodicity that increases when approaching the linear threshold ($T\gtrsim 10^3$), which is qualitatively different from Dimits shift dynamics.}
This recalls the observations in \cite{Hoffmann2023GyrokineticShift}, where the Dimits shift is not observed when considering $d_{\max}=2$, suggesting that the HEL closure scheme may not be sufficient to compensate for the absence of higher-order kinetic effects.
Since the Dimits shift results from the formation of ZFs, this suggests that the HEGS lacks mechanisms favorable to ZF formation.
This observation provides further evidence that these mechanisms are embedded in the higher-order GMs, particularly in the parallel and perpendicular heat fluxes moments, $q_\parallel$ and $q_\perp$, and the energy-weighted pressure tensor \citep{Beer1995} related to GMs such that $p>2$ and $j>1$.
At the same time, our results also suggest that a higher-order gyrofluid \modifi{system of equations} may be able to reproduce the Dimits shift in the tokamak geometry, as it would include the higher-order moments that are responsible for ZF formation.

\begin{figure}
    \centering
    \includegraphics[width=0.5\linewidth]{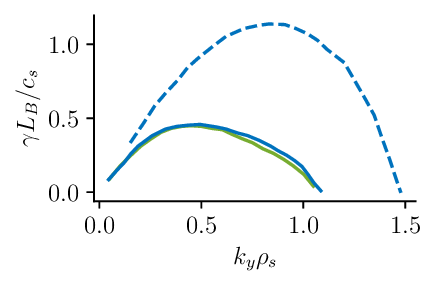}
    \caption{\modifi{Linear growth rates of ITG simulations in the Z-pinch geometry with the GK simulations (solid) and HEGS (dashed). The hybrid geometry (green) is obtained with the $s-\alpha$ geometry setting $q_0=100$, $\epsilon=0.001$ and $\hat s =0$.}}
    \label{ch6_fig:lin_ZP}
\end{figure}
\begin{figure}
    \centering
    \includegraphics[width=0.8\linewidth]{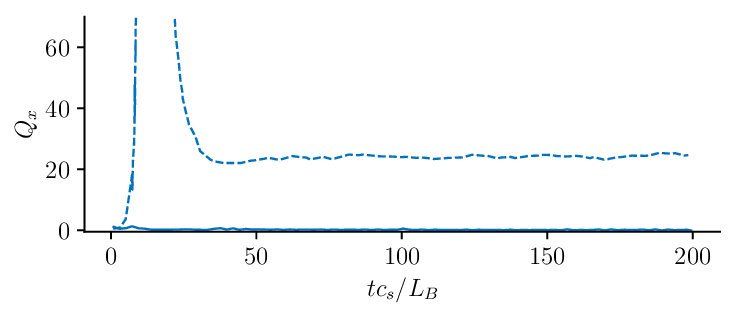}
    \caption{\modifi{Time traces of the radial heat flux in the Z-pinch geometry with the GK simulations (solid) and HEGS (dashed).}}
    \label{ch6_fig:HF_time_traces_ZP}
\end{figure}
We can now leverage the capability of \gyacomo to perform \modifi{GK simulations} in both tokamak and Z-pinch geometries to isolate the impact of the geometry from the kinetic effects missing in the HEL limit.
We consider the same parameters as in the tokamak case and compare the Z-pinch linear and nonlinear results in Figs. \ref{ch6_fig:lin_ZP} and \ref{ch6_fig:HF_time_traces_ZP}.
The discrepancies between the \modifi{GK simulations and the HEGS} observed in the linear growth rates are of the same nature as those observed in the tokamak case, where the unstable mode amplitude and spectrum are increased when considering the HEGS.
In the nonlinear case, both \modifi{GK and HEGS} yield a ZF dominated system. The \modifi{GK simulations} predict a suppression of the transport, whereas the HEGS allows for a finite transport level, which can be attributed to the $k_\parallel$ turbulence observed \modifi{in Sec. \ref{sec:nonlinear_results}}.
The conclusion from this experiment is threefold. 
First, it points out that the discrepancies between the \modifi{HEGS and GK simulations} are not solely due to the geometry.
Second, it indicates that weak turbulence regimes are harder to capture in the HEL than strongly turbulent regimes.
Third, it demonstrates that the Z-pinch geometry is more favorable to ZF formation than the $s-\alpha$ geometry, regardless of the model used, as the Z-pinch simulations show a lower transport level despite larger linear growth rates.

\modifi{
To interpret this geometry dependence, we propose a simplified physical picture based on the role of curvature uniformity in ZF generation. 
In the framework of plasma instability theory \citep{Rogers2005,Ricci2006a}, ZFs are driven by a secondary instability arising from the nonlinear interaction of primary ITG modes. The dominant ZF radial wavenumber $k_x^{\mathrm{ZF}}$ is thus set by the primary instability spectrum.
In the Z-pinch geometry, the curvature is uniformly unfavorable along the field line, yielding a $z$-independent primary growth rate spectrum, $\partial_z \gamma_{\mathrm{ITG}} = 0$. 
This parallel coherence promotes a secondary instability that is uniform in the parallel direction, which drives a single dominant ZF mode across the entire domain. As a result, the energy extracted from the primary instability is channeled efficiently into one coherent ZF structure that can grow to a large amplitude and effectively suppress turbulence.
In contrast, the tokamak geometry features a $z$-dependent curvature that modulates the local ITG drive along the field line, $\partial_z \gamma_{\mathrm{ITG}} \neq 0$.
Different poloidal locations may therefore favor different ZF radial scales, $k_x^{\mathrm{ZF}} = k_x^{\mathrm{ZF}}(z)$, leading to potentially destructive interference among competing zonal structures and ultimately reducing the ZF amplitude and turbulence suppression efficiency.
Since the tokamak geometry reduces to a Z-pinch geometry in the $\epsilon \to 0$, $q_0 \to \infty$ limit (see Fig.~\ref{ch6_fig:lin_ZP}), increasing the aspect ratio or the safety factor could lead to an improvement of the confinement.
It is worth noting that this simplified picture neglects several potentially important effects: (i) three dimensional instabilities, (ii) the role of magnetic shear in modulating ZF coherence along the field line, and (iii) tertiary instabilities that may limit the ZF amplitude, particularly in the Z-pinch geometry, where stronger ZF shear makes them more susceptible to breakdown.
A quantitative assessment of these mechanisms is left for future work.
}

\section{Conclusions}
\label{sec:conclusions}
We study an HEL asymptotic closure of the GM hierarchy, establishing a pathway from a GK formulation to a reduced fluid representation. By expanding the Hermite-Laguerre gyroaveraging kernels in $\tau = T_i/T_e$ and retaining the minimal $\mathcal{O}(\tau)$ contributions required for consistency, we derive the HEL--GM system and demonstrate its analytical equivalence with the Ivanov Z-pinch fluid model with an empirically calibrated collisionality parameter. This derivation implies that the Ivanov model is an analytical limit of the GM approach, thus opening a new route to deriving reduced models.

Our numerical simulations with the \gyacomo code yield several principal results. 
Closure verification in linear Z-pinch simulations confirms that four retained GMs, corresponding to density, parallel velocity, and parallel and perpendicular temperatures, form a closed set in the $\tau\ll 1$ limit.
The introduction of an empirical constant factor ($c_f=4$) is sufficient to reconcile our Dougherty collision model with the published Landau-based operator \citep{Ivanov2020ZonallyTurbulence}. This result indicates that most of the kinetic effects captured in the Landau operator, such as the velocity space dependence of the collision frequency, are lost when considering the $\tau\ll1$ limit.
Previous nonlinear simulation results of Z-pinch turbulence are reproduced. In particular, HEGS retrieve the heat-flux levels quantitatively and the bursty or blow-up behavior at low collisionality, capturing the transition where ZFs weaken.
Parallel domain elongation studies yield asymptotic transport plateaus consistent with previous analyses \citep{Ivanov2022DimitsTurbulence,Volcokas2023UltraTokamaks}.

Extending the HEL to the tokamak $s{-}\alpha$ geometry, we compare its results with $\tau=1$ GK simulations. The HEGS overpredict linear growth rates and spectral broadening, yet preserves the qualitative heat-flux temporal structure. At the same time, the HEGS show a reduced or absent Dimits shift, indicating that higher-order moments (parallel and perpendicular heat fluxes and pressure-tensor components) have a crucial role in zonal-flow amplification in tokamak configurations and are not recoverable within the lowest-order HEL truncation. 

Finally, the impact of geometry on ZF formation is assessed.
The Z-pinch geometry presents stronger ZF mitigation of transport with respect to the CBC tokamak geometry. This effect can be linked to the Z-pinch bad curvature, which allows coherent ZF layers to span the entire domain and persist, hence suppressing turbulence more effectively despite higher linear growth rates. In contrast, the tokamak geometry's varying curvature along field lines induces competing zonal modes, which can disrupt ZF coherence and weaken their regulatory effect on turbulence.

The findings presented here underline the physical effects retained and those lost under the HEL reduction. The model retains the turbulence drive mechanisms, the ZF saturation mechanism in a uniform-curvature geometry such as that of the Z-pinch, and the parallel decorrelation effects governing 3D saturation in a turbulence-dominated regime.
However, the absence of higher-order kinetic moments, including parallel and perpendicular heat fluxes and pressure-tensor components, prevents the HEL model from accurately reproducing phenomena such as the Dimits shift in tokamak geometries.

Applying the $\tau\ll1$ limit to the GM hierarchy offers a new closure scheme that, at first order, is able to qualitatively reproduce transport in turbulence-dominated regimes for both Z-pinch and tokamak geometries, even when the hot-electron assumption is violated.
It is now possible to systematically extend this closure scheme to higher-order moments by retaining higher-order $\tau$ contributions and higher-order GM equations. 
Our results suggest that the resulting higher-order fluid model should be able to capture the Dimits shift in the tokamak geometry, extending the range of applicability of the HEL-GM model.

\section*{Acknowledgements}
The authors gratefully acknowledge helpful discussions with A. Vol\v{c}okas, S. Brunner, J. Ball, and T. Adkins.
The simulations presented herein were carried out in part on the CINECA Marconi supercomputer under Project TSVVT422 and in part at the CSCS (Swiss National Supercomputing Center). This work was carried out within the framework of the EUROfusion Consortium, via the Euratom Research and Training Programme (Grant Agreement No. 101052200—EUROfusion), and was funded by the Swiss State Secretariat for Education, Research and Innovation (SERI). The views and opinions expressed herein are, however, those of the author(s) only and do not necessarily reflect those of the European Union, the European Commission, or SERI. Neither the European Union, the European Commission, nor SERI can be held responsible for them.

\newpage

\bibliographystyle{jpp}
\bibliography{references}
\end{document}